\documentclass{elsart}
\usepackage{graphicx}

\usepackage{amssymb}

\begin{document}

\begin{frontmatter}



\title{The Hadron Hose:  Continuous Toroidal Focusing for Conventional Neutrino Beams}

 \author[fermi]{J. Hylen}
 \author[fermi]{D. Bogert}
 \author[fermi]{R. Ducar}
 \author[ihep] {V. Garkusha}
 \author[ut]{J. Hall}
 \author[fermi]{C. Jensen}
 \author[ut]{S.E. Kopp\corauthref{myemail}}
 \corauth[myemail]{Corresponding author e-mail kopp@mail.hep.utexas.edu}
 \author[ut]{M. Kostin}
 \author[ut]{A. Lyukov}
 \author[fermi]{A. Marchionni}
 \author[fermi]{M. May}
 \author[harvard]{M.D. Messier}
 \author[tufts] {R. Milburn}
 \author[ihep] {F. Novoskoltsev}
 \author[ut]{M. Proga}
 \author[fermi]{D. Pushka}
 \author[fermi]{W. Smart}
 \author[fermi]{J. Walton}
 \author[ihep] {V. Zarucheisky}
 \author[ut]{R.M. Zwaska}
 \address[fermi]{Fermi National Accelerator Laboratory, Batavia, Illinois  60510  U.S.A.}
 \address[harvard]{Dept. of Physics, Harvard University, Cambridge, Massachussettes  02138  U.S.A.}
 \address[ut]{Dept. of Physics, University of Texas, Austin, Texas  78712  U.S.A.}
 \address[ihep] {Institute for High Energy Physics, Protvino, Russia}
 \address[tufts] {Tufts University, Medford, MA  02155  U.S.A.}

\vskip -.25 cm

\begin{abstract}

We have developed a new focusing system for conventional neutrino beams. The ``Hadron Hose'' is a 
wire located in the meson decay volume, downstream of the target and focusing horns.  The wire is pulsed
with high current to provide a toroidal magnetic field which continuously focuses mesons.  
The hose increases the neutrino event rate and reduces differences between near-field and far-field 
neutrino spectra for oscillation experiments.  We have studied this device as 
part of the development of the Neutrinos at the Main Injector (NuMI) project, but it might also be 
of use for other conventional neutrino beams.
\end{abstract}

\end{frontmatter}



\section{Introduction}

\begin{figure}[t]
\centering
\includegraphics*[width=140mm]{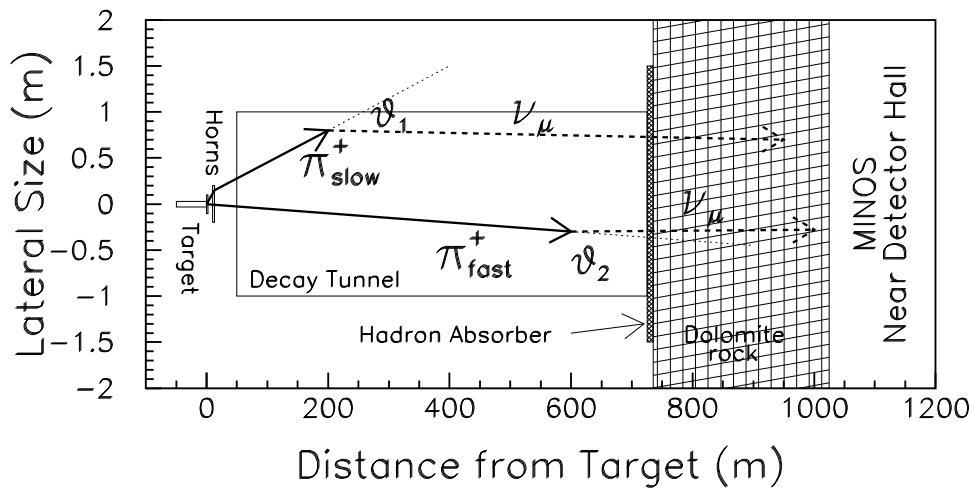}
\caption{Schematic layout of the NuMI neutrino beamline, showing the locations of the target, focusing horns, 
decay volume, and steel and earth shielding in front of the near MINOS detector.  Two sample pion trajectories,
are shown.  Soft pions tend to enter the decay volume at large angles, while stiff pions enter at smaller angles.}
\label{numinohose}
\end{figure}

The Hadron Hose is a current-carrying wire used to instrument the decay volume of a conventional
neutrino beam with a toroidal focusing field for a sign-selected meson beam.  Conventional neutrino
beams are tertiary beams, with the primary proton beam producing $\pi$ and $K$ meson secondaries in
a target.  Beamline elements downstream of the target, called horns \cite{horn}, produce a toroidal magnetic
field to sign- and momentum- select the mesons, focusing them toward an evacuated or 
Helium-filled volume, where the mesons decay into the tertiary neutrino beam.  Steel
and earth shielding absorb remnant protons, mesons, and muons at the end of 
the decay volume (see Figure~\ref{numinohose}).

\begin{figure}[t]
\centering
\includegraphics*[width=60mm]{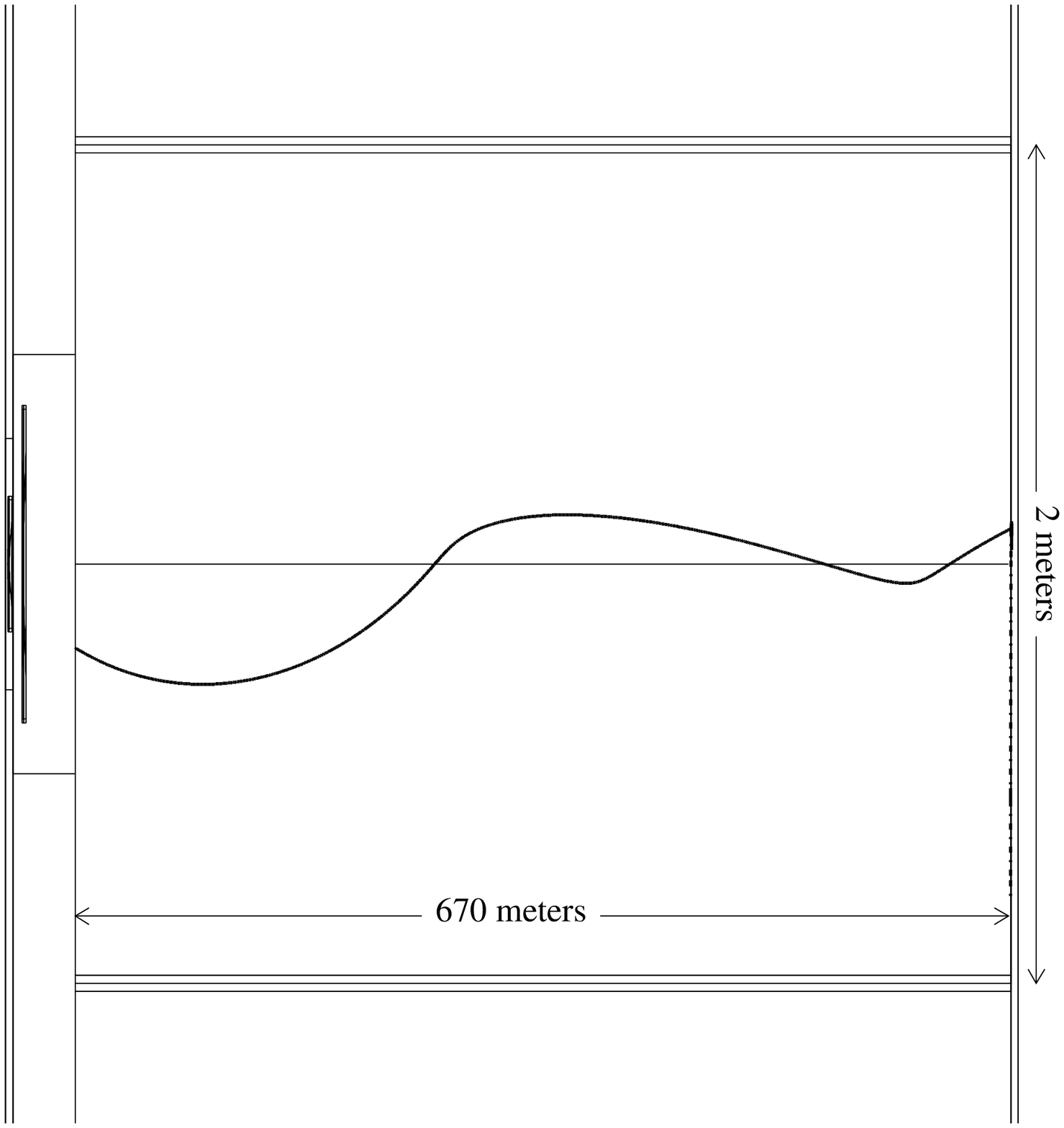}
\hfill
\includegraphics*[width=60mm]{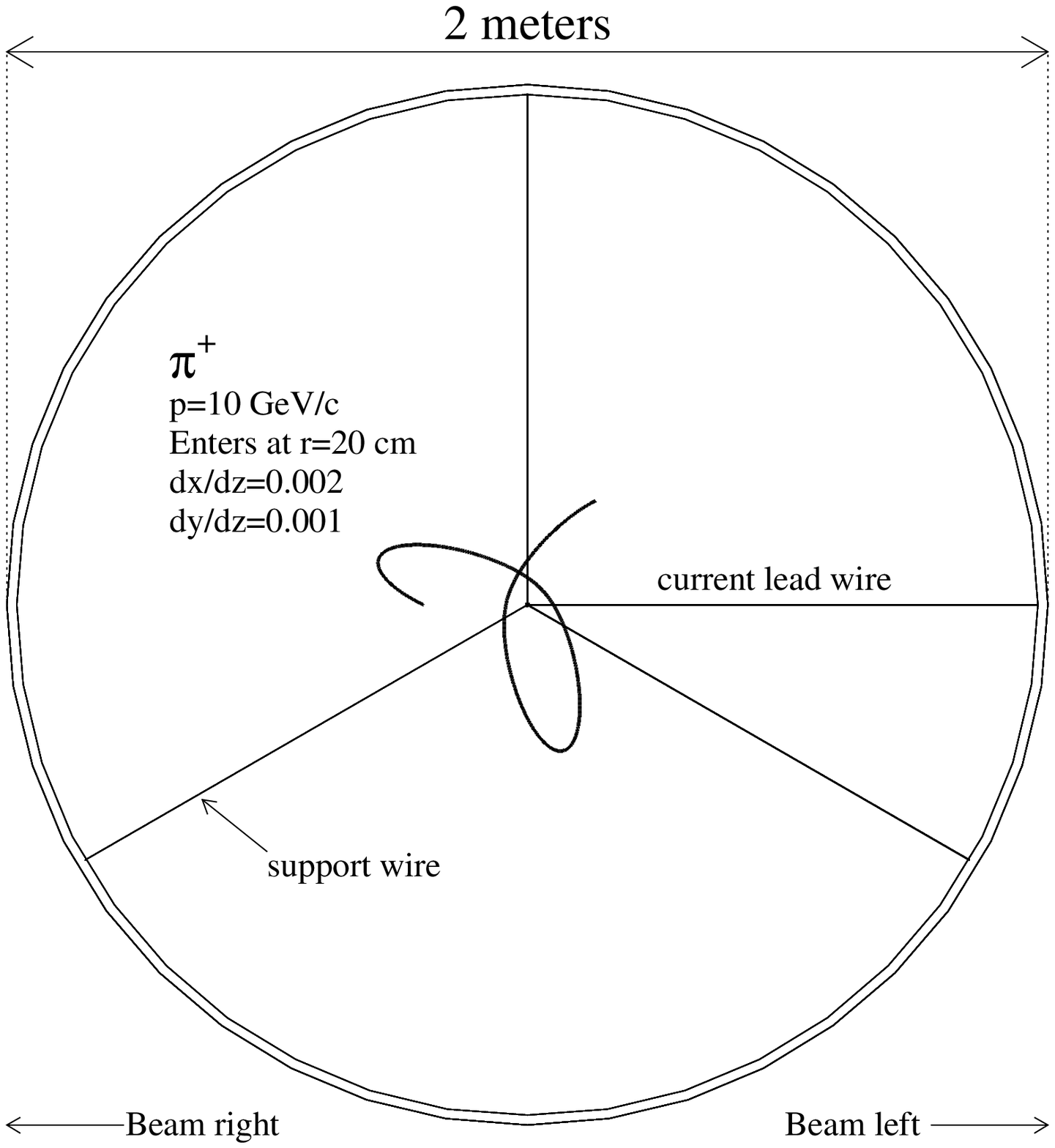}
\caption{Sample orbit of a 10~GeV pion in the NuMI beamline including the hadron hose.  The toroidal field of the hose
wire captures the pion in orbit around the wire.  The orbit randomizes the decay angle between the pion direction and the
direction of neutrinos that reach the MINOS detectors.}
\label{numihose}
\end{figure}

In conventional neutrino beams to date \cite{lederman,oldcern,wanf,bnl,k2k,ferminubeam,ferminubeam2}
the mesons freely propagate through the decay volume.\footnote{In some of the beamlines, the
mesons are first focused by quadrupole or horn magnets located just after the production target and before
the decay volume, while other
beams have been 'bare target' beams where the mesons from the target directly enter the decay volume without focusing.}
The toroidal field of the Hadron Hose provides continuous focusing of the secondary meson beam throughout the 
decay volume.  Mesons spiral around the hose wire and are drawn away from the decay volume walls where they 
might interact before decaying.  A pion orbit in the NuMI beamline with the Hadron Hose is shown in 
Figure~\ref{numihose}.  The spiraling motion randomizes 
the decay angles of the mesons, which reduces the need to model detector and beamline acceptances in Monte Carlo
calculations of the expected neutrino fluxes at the detectors.

Toroidal magnetic fields have been used for plasma beams \cite{plasma1,plasma2} and the possibility of
using toroidal fields to focus charged particle beams was noted 
by Van~der~Meer \cite{vandermeer,regenstreif}.  In this article,
we study the utility of such a focusing system for neutrino beams and the technical feasibility of placing such
a wire in an evacuated region with high particle fluences.  While we have developed this device for the 
Neutrinos at the Main Injector (NuMI) project, such a system might be of utility at future neutrino facilities
such as the JHF \cite{jhf} or conventional neutrino ``super beams'' \cite{superbeam1,superbeam2,superbeam3}.

This article procedes as follows:  We discuss the focusing properties of the hose in 
Sections~\ref{numi}~-~\ref{hose}.  Sections~\ref{requirements}-\ref{elecdesign} present the mechanical 
and electrical design which was proposed for NuMI.  Sections~\ref{breakdown} through \ref{radiological} discuss some of 
the hardware studies to demonstrate that the hose wire can survive in the radiation field of a neutrino beamline 
without significant failure of wire segments.  Section~\ref{conclude} concludes.

\section{NuMI Beamline}
\label{numi}

For NuMI, 120~GeV protons will be extracted from the Main Injector \cite{NuMI} and focused downward by 58~mRad, 
to strike a 0.94~m long graphite target.  The bunch length is $8.6~\mu$sec, and the cycle time $1.87$~sec. 
The beamline is designed for $3.7\times 10^{13}$ protons/pulse and $3.8\times 10^{20}$ protons/year.

\begin{figure}[t]
\includegraphics[width=70mm]{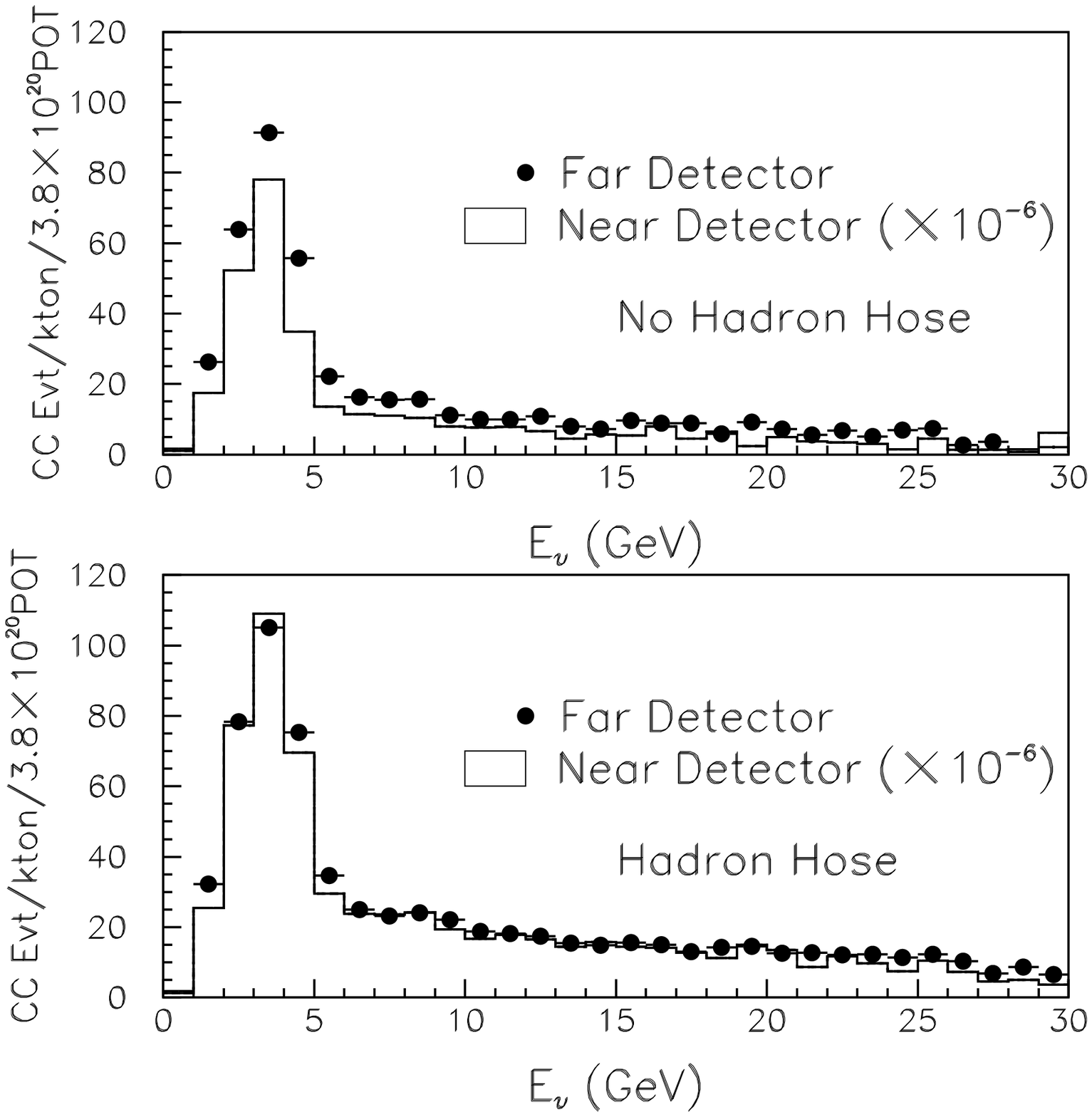}
\hfill
\includegraphics[width=70mm]{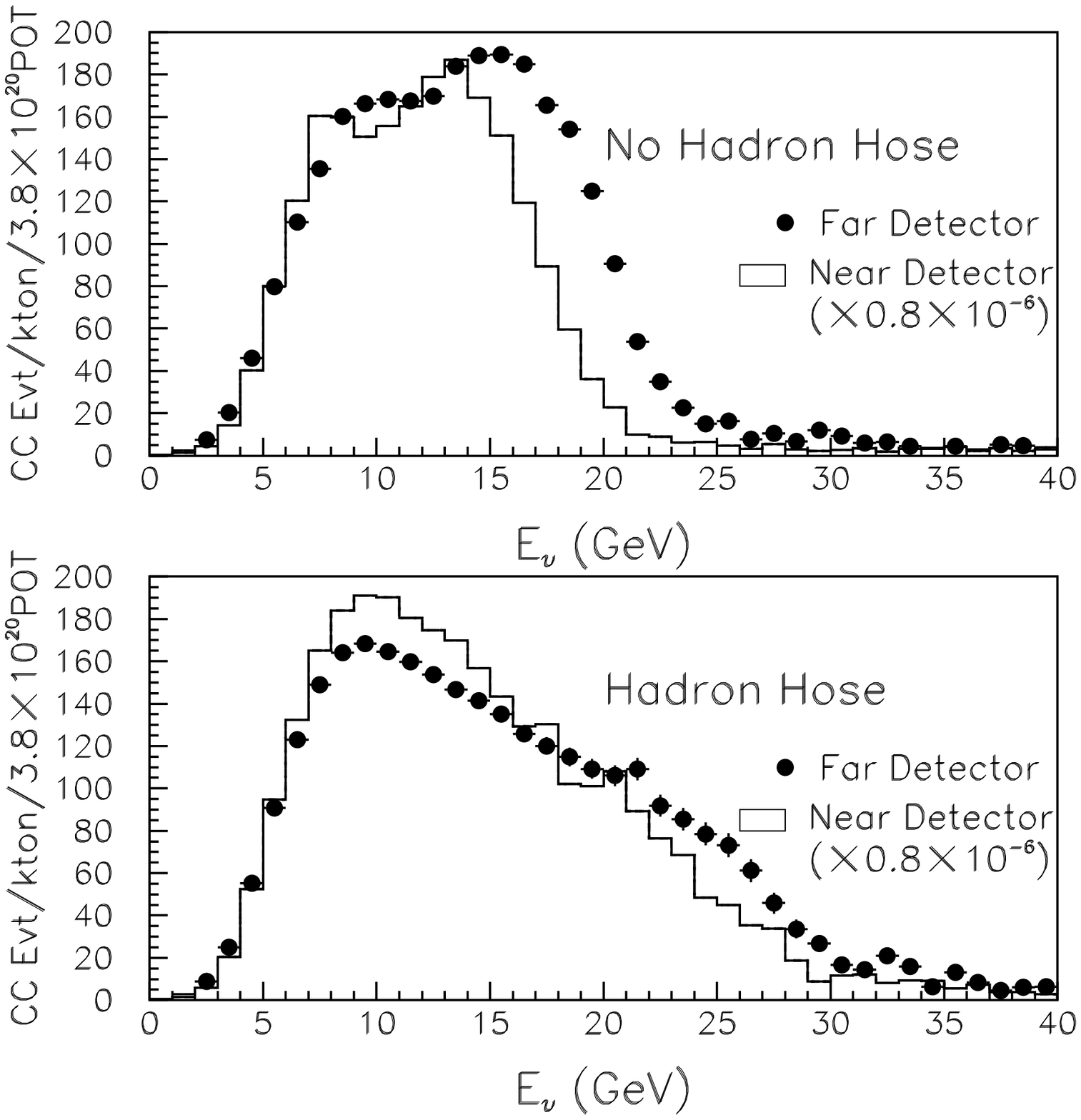}
\caption{Neutrino energy spectra in the far and the near ($\times10^{-6}$) MINOS detectors,
computed without (above) and with (below) the Hadron Hose.  The left plots are for the NuMI
low-energy beam and the right are for the high energy beam.  }
\label{lespectra}
\end{figure}

NuMI will have two focusing horns \cite{NuMI} pulsed at 200~kA after the target.  The secondary hadrons 
enter a 675 m long, 1~m radius evacuated decay volume.  MINOS \cite{MINOS} is a 2-detector neutrino experiment.  
A 980~ton near detector measures the neutrino energy spectrum and rate produced at Fermilab.
A 5400~ton far detector is located in the Soudan mine in Minnesota, 735~km from Fermilab.   The NuMI beam can
be adapted to produce a low ($E_{\nu}^{peak}\sim3$~GeV), medium ($E_{\nu}^{peak}\sim6$~GeV), or high
($E_{\nu}^{peak}\sim12$~GeV) energy neutrino event spectrum.
Figure~\ref{lespectra} shows the expected spectra in the two detectors for
the low-energy and high-energy beam options assuming no new physics resulting in neutrino disappearance.
In Figure~\ref{lespectra}, the vertical axis is the number of expected 
charged current neutrino interactions expected per kiloton of detector mass 
per $4\times10^{20}$~protons on target ($\sim$1~NuMI~year).

The neutrinos in the peaks of the spectra of Figure~\ref{lespectra} 
come from pions focused by the
horns, whereas the neutrinos in the high energy tail come from poorly focused pions (pions that
travel through the necks of the horns).  
The high energy tail is nonetheless important to the experiment because 
it provides us a control sample to demonstrate a region without oscillation effects.\footnote{In
the MINOS experiment, a $\nu_{\mu} \rightarrow \nu_X$ oscillation with $\Delta m^2~=~3-5\times10^{-3}$~eV$^2$
results in a depletion of neutrinos centered around 1.8-3.0~GeV, while the spectrum above 10~GeV is unchanged.}
It is important to make the best prediction of the far-field energy spectrum to search for any kind of 
spectral distortions.

It is desirable to have the neutrino energy spectra at the two detector sites as similar as possible so that
the near detector site strongly constrains the calculation of the spectrum at the far site.
Any relative difference
observed at the far detector is evidence for new physics such as neutrino oscillations.  In practice,
the two detector spectra are not identical: the vastly different solid angles subtended by the two 
detectors results in an energy-dependent acceptance difference.  This difference is evident in the 
neutrino spectra of Figure~\ref{lespectra}.

The acceptance difference between the two detectors is accounted
for in the ``far-to-near ratio'', which is the factor by which the near detector spectrum must be multiplied
to predict the far detector spectrum. It is defined as:
\begin{equation}
N_{\mbox{far}}^i =\mathcal{R}_{FN} N_{\mbox{near}}^i.
\end{equation}
where $N_{\mbox{near}}^i$ is the observed number of events in the $i^{\mbox{th}}$ energy bin in the near
detector, and $N_{\mbox{far}}^i$ is the predicted number of events in the $i^{\mbox{th}}$ bin in the far 
in the absence of new physics.  Uncertainties in the calculation of $\mathcal{R}_{FN}$
lead to systematic uncertainties in the predicted far detector spectrum which limit the ultimate reach of 
physics searches.

The calculation of $\mathcal{R}_{FN}$ is complicated by the extended length 
and finite aperture of the neutrino beam geometry.  If the neutrinos were produced from a point source, 
then $\mathcal{R}_{FN}$ could be estimated by $\mathcal{R}_{FN} = Z_{near}^2/Z_{far}^2$,
where $Z_{near}$ ($Z_{far}$) is the distance from the target to the near (far) neutrino detector.
Considering that neutrino beamlines are an extended source,
one could weight this extrapolation factor by the pion lifetime along the length of the decay tunnel:
\begin{equation}
\mathcal{R}_{FN} = {
{\int_{z = 0 m}^{z = 725 m}  e^{-\frac{0.43m_{\pi}z}{E_{\nu}c\tau}} {1 \over (Z_{far}-z)^2} dz}
\over
{\int_{z = 0 m}^{z = 725 m}  e^{-\frac{0.43m_{\pi}z}{E_{\nu}c\tau}} {1 \over (Z_{near}-z)^2} dz} }
\label{rfneqn}
\end{equation}
where the integral is over the length of the decay tunnel (725~m in the case of the NuMI beamline) 
and the substitution $E_{\pi}\sim E_{\nu}/0.43$ has been made.

\begin{figure}[t]
\includegraphics[width=70mm]{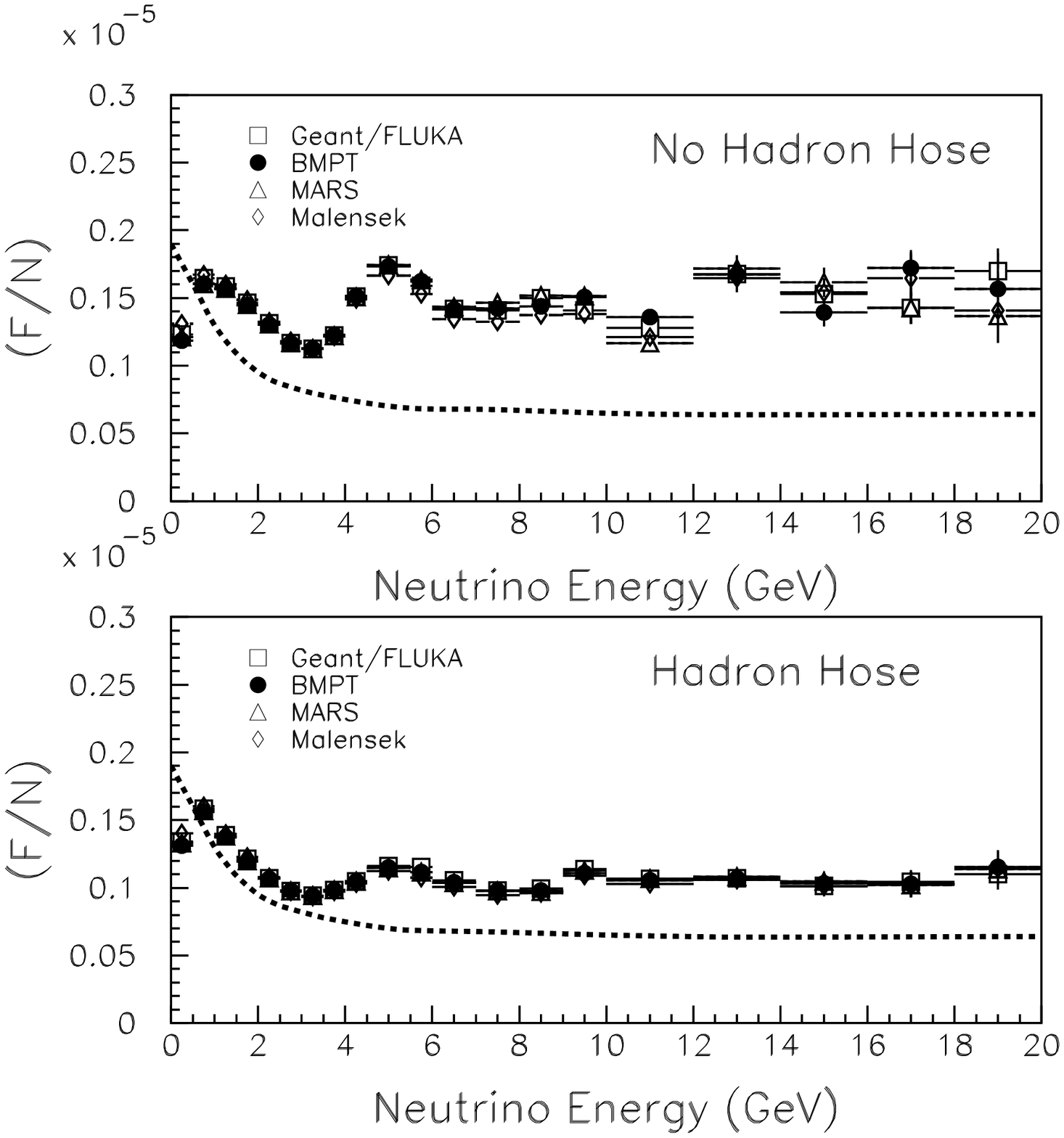}
\hfill
\includegraphics[width=70mm]{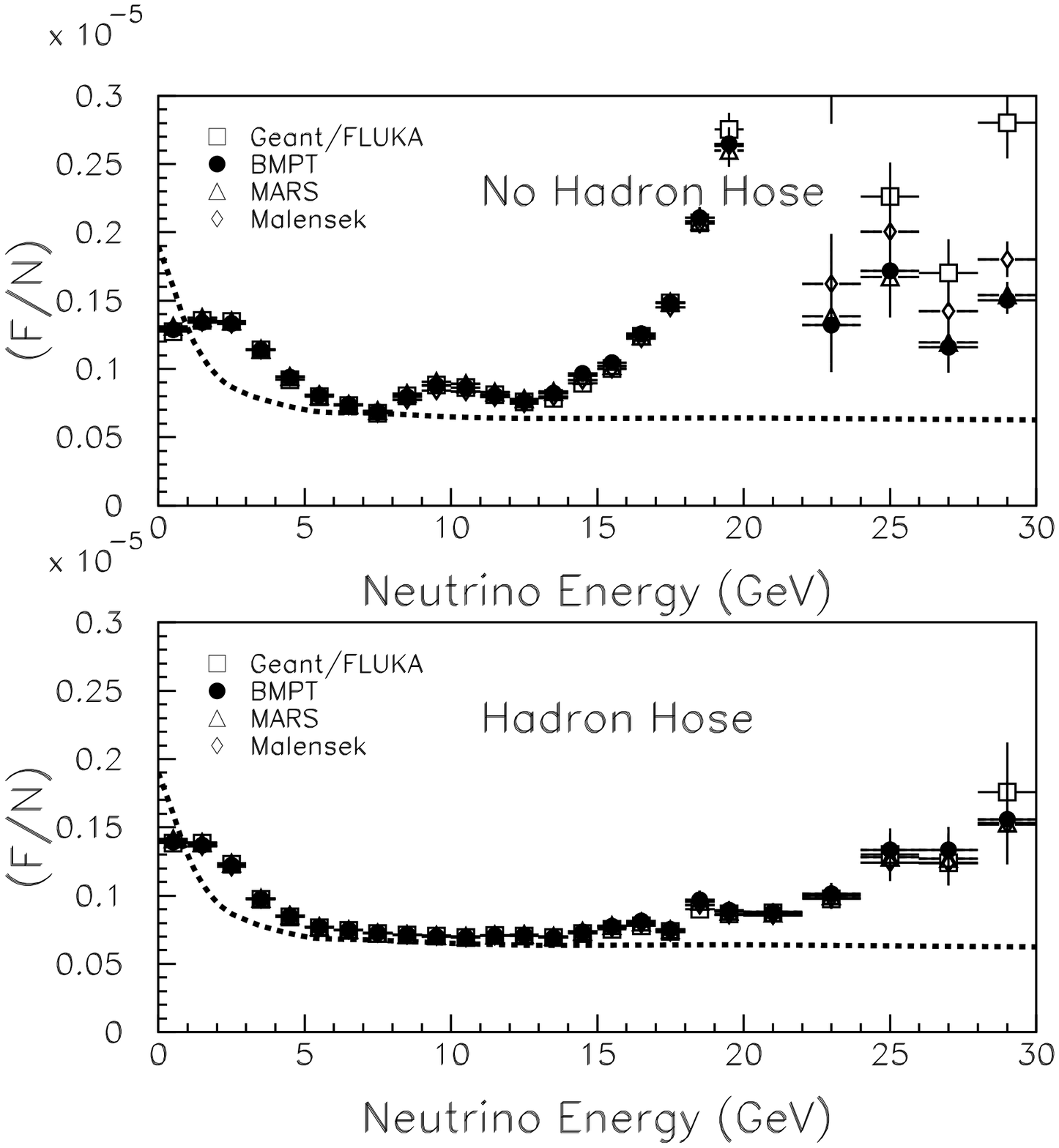}
\caption{The Far-over-Near ratio (F/N) calculated for the NuMI low energy beam using several predictions
\cite{BMPT,MARS,Malensek,GFLUKA,fluka} for the production of hadrons in the NuMI target.  Shown is the prediction
without (upper plot) and with (lower plot) the hadron hose in the low-energy (left) and high-energy
(right) beams.  The dotted line is the result of 
the simplified calculation of the far-over-near ratio from Equation~\ref{rfneqn}.}
\label{modelspread}
\end{figure}

However, even this estimate of the near-far extrapolation is a simplification:  not all 
decaying pions produce neutrinos within the finite acceptances of the two detectors and not all pions are 
able to decay before interacting along the decay pipe walls.  Furthermore, the correct relation between the 
pion and neutrino energy is 
\begin{equation}
E_{\nu} = \frac{0.43 E_{\pi}}{1+\gamma^2\theta^2}
\label{enrgycorrel}
\end{equation} 
where $\gamma$ is the pion relativistic boost and
$\theta$ is the decay angle of the pion.  In fact, because low-momentum pions tend to enter the decay volume
at wider angles and decay far upstream in the decay
volume, while high-momentum pions tend to propagate much further, the acceptance of the near detector for
high momentum neutrinos is larger than for low-momentum neutrinos.  The acceptance correction $\mathcal{R}$$_{FN}$
given in Equation~\ref{rfneqn} is incorrect in an energy-dependent manner because of the correlation 
in Equation~\ref{enrgycorrel}.  A more detailed extrapolation of the measured near spectrum 
to the far detector requires a Monte Carlo calculation which tracks secondary hadrons produced in the 
target through the beamline optics.  A comparison of the naive calculation of Equation~\ref{rfneqn} 
to a {\tt GEANT} \cite{GFLUKA} simulation of the NuMI beamline is shown in Figure~\ref{modelspread}.

\section{Hadron Hose}
\label{hose}

The Hadron Hose consists of a wire at the center of the decay
tunnel and carries a 1000~A peak current pulse.  The current provides a toroidal
magnetic field which continually focuses positive particles back toward the center of the decay pipe.  
A typical meson executes 2-3 orbits around the wire over the length of the decay pipe.

\begin{figure}[t]
\centering
\includegraphics[width=60mm]{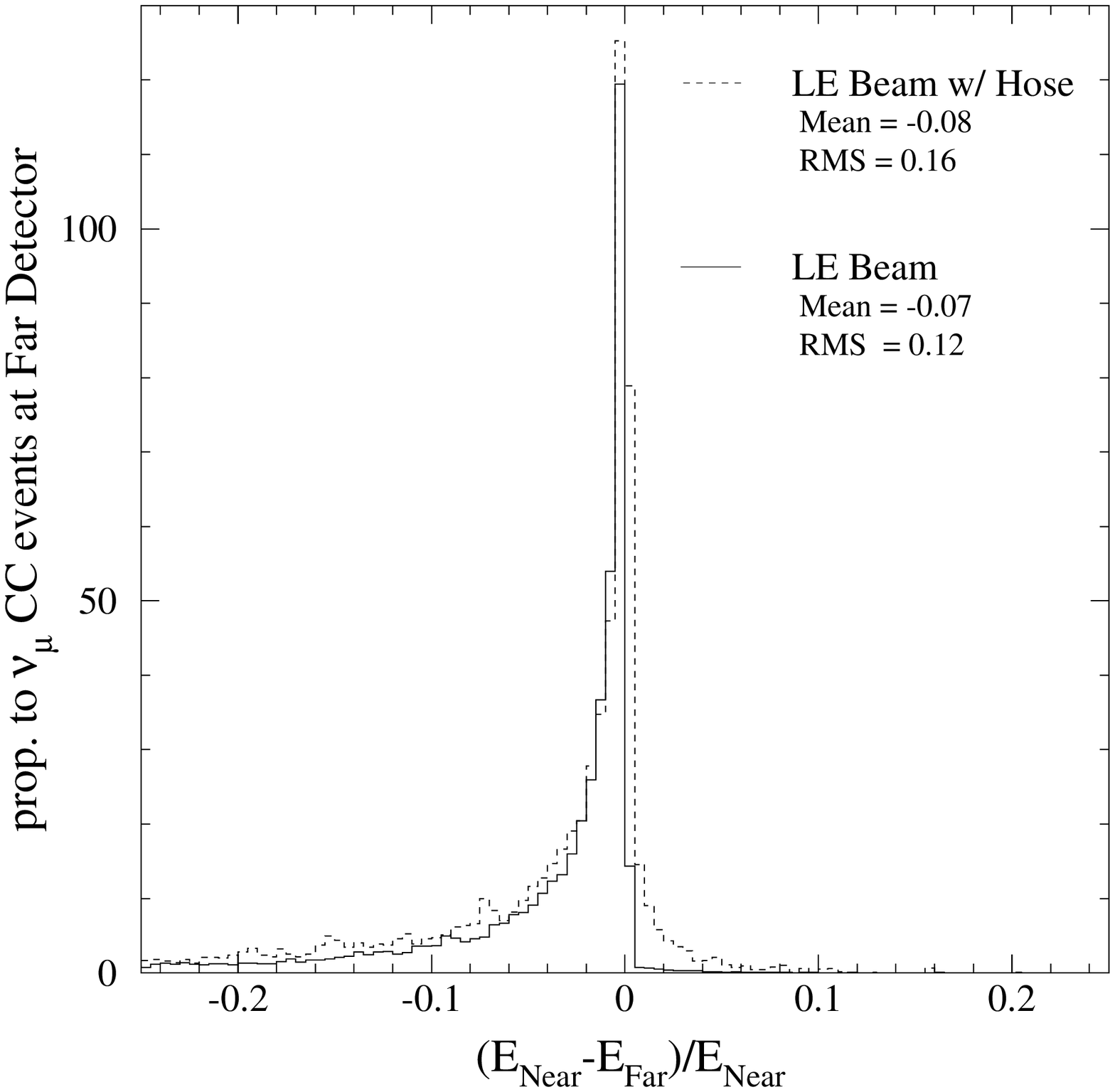}
\includegraphics[width=60mm]{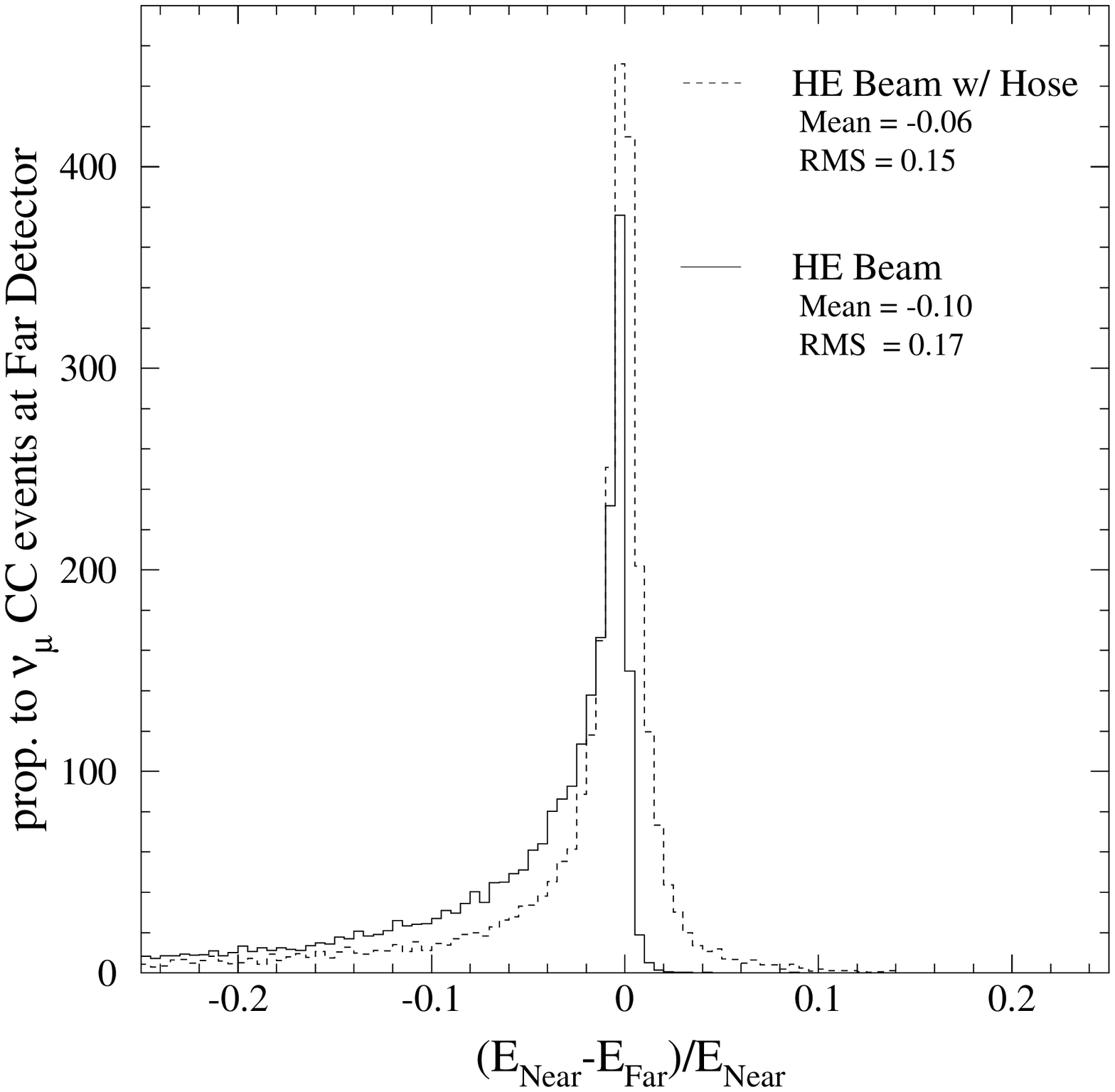}
\caption{The difference in the energy of the neutrino that a given pion would emit toward the far detector
to that it would emit toward the near detector in NuMI/MINOS (see text) in the low-energy (left) and high-energy (right) beams,
normalized to the near detector energy.}
\label{nearfarangles}
\end{figure}

The hadron hose contributes three essential features.  First,  $\pi$'s and $K$'s that otherwise diverge out to the 
decay pipe walls and interact before they decay are given a restoring force back to the decay pipe center.
Thus, $\pi$'s/$K$'s travel farther and have a greater chance to decay, so the neutrino 
event rate in the far detector is increased (see Table~\ref{fluxes}).  Second, the pion decay distribution
along the beamline direction more nearly follows a simple exponential, which reduces the difference in 
acceptance of the near detector to low-momentum {\it vs.} high-momentum pions.  Thus, the extrapolation
factor $\mathcal{R}$$_{FN}$ more closely follows that given by the pion lifetime in Equation~\ref{rfneqn}, as 
can be seen in Figure~\ref{modelspread}.
Third, the pion orbits effectively randomize the decay angle between the pion direction and the neutrino that 
hits the MINOS near and far detectors.  This randomization
washes out the kinematic correlation between neutrino energy and decay angle in Equation~\ref{enrgycorrel},
which otherwise always produces a softer spectrum in the near detector than the far detector.  The difference
is especially visible at the high energy edge of the high energy beam in Figure~\ref{lespectra}.  
The randomization is a larger effect for the high energy beam because the Lorentz boost at high energy otherwise 
increases the sensitivity to the decay angle in Equation~\ref{rfneqn}.   The effect is also demonstrated in 
Figure~\ref{nearfarangles}, which plots the ratio of neutrino energies in the near and far detector that would
arise from the decay of a given pion in the low-energy or high-energy NuMI beam configurations.

\begin{table}[b]
\begin{center}
\begin{tabular}{|l|c|c|}
\hline
\textbf{Beam} & \textbf{Events in peak} & \textbf{Overall Events} \\ \hline
LE     & 261 & 474 \\
LE-HH  & 327 & 732 \\
HE     & 2694 & 2745 \\ 
HE-HH  & 2870 & 2983 \\ 
\hline
\end{tabular}
\caption{Charged current event yields in the far MINOS detector in the low 
and high energy
beam configurations, with and without Hadron Hose.  The peak refers to the regions
$E_{\nu}<$6 and 30 GeV, for low and high energy beams, while overall refers to $E_{\nu}<40$ GeV. 
The units are events per kiloton of far detector mass per $3.8 \times 10^{20}$~protons on target.}
\label{fluxes}
\end{center}
\end{table}

By minimizing acceptance effects of the detectors to the beam,
the hose reduces the experiment's sensitivity to input data to the Monte Carlo.  The largest
systematic uncertainty in $\mathcal{R}_{FN}$ for NuMI are the cross sections 
for production of $\pi$'s/$K$'s in the target, $d^2\sigma/dx_F dp_T$.  These have been measured 
\cite{SPY,Atherton,Barton,Brenner,Eichten}, but gaps in these data sets
in the $x_F$ and $p_T$ range relevant for NuMI, 10-20\% disagreements between data sets, and systematic 
uncertainties in scaling these invariant cross sections to the NuMI beam energy and target material 
complicate efforts to model or parameterize the data for all $x_F$ and $p_T$.  Using several
parameterizations \cite{BMPT,MARS,Malensek,GFLUKA,fluka}, 20-30\% variations are found in neutrino flux predictions
for the near and far detectors, and more importantly, up to 5\% variations are predicted in the F/N ratio 
(see Figure~\ref{modelspread}).  These variations are reduced with hose focusing.

Besides minimizing the experiment's sensitivity to variations in pion productions cross sections,
the Hadron Hose loosens accuracy criteria for other beamline components.  For NuMI, the potential benefits include
increasing the allowed eccentricity of the
inner conductor of the first focusing horn from $<$0.08~mm at its neck position to $<$0.12~mm if hose focusing is
implemented.  In addition,
the spatial alignment of the first horn transverse to the beamline is relaxed from $\pm0.8$~mm to $\pm1.0$~mm.
Relative current variations between the two horns can be as large as $\pm1.5$\% (c.f. $\pm1.0$\%).  
Finally, the alignment of the NuMI decay pipe, previously required to 
be within 2~cm all along its 675~m length, would be relaxed to 3~cm.

\begin{figure}[t]
\centering
\includegraphics[width=100mm]{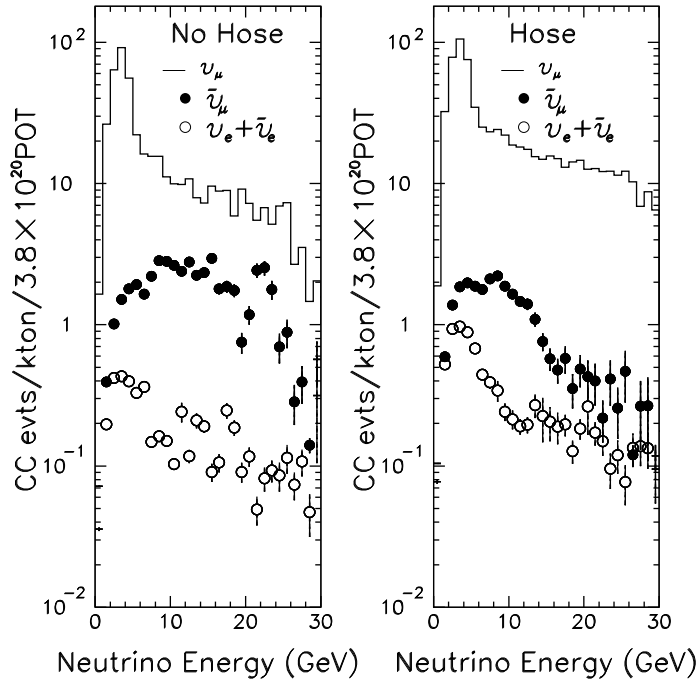}
\caption{The composition of neutrino species in the NuMI beam, with and without the hadron hose in the
NuMI low-energy beam.}
\label{contamination}
\end{figure}

The Hadron Hose alters the backgrounds from $\overline{\nu}_{\mu}$, $\nu_{e}$, and 
$\overline{\nu}_{e}$ in the beam, as shown in Figure~\ref{contamination}.  
The $\overline \nu_{\mu}$ backgrounds, which come from $\pi^- \rightarrow \mu^- \overline{\nu}_{\mu}$ decays,
are reduced because the hose field defocuses $\pi^-$.
The $\nu_{e}$ and $\overline{\nu}_{e}$ backgrounds are enhanced because
$\mu^+$ daughters from $\pi^+$ decay are also focused by the hose field.
In the case of NuMI, the factor of 2 increase in $\nu_{e}$ does not have a serious impact on $\nu_{\mu} \rightarrow
\nu_{e}$ searches, since the putative signal would also go up by a factor of 30\%, so 
signal/$\sqrt{\mbox{background}}$ is unchanged.  For other experiments, the increase might be less 
because the NuMI decay pipe is quite long.

A Hadron Hose could also be designed to operate at currents of
1.5, 2.0, or even 4.0~kA.  Such a design would be more ambitious
than the NuMI proposal, requiring faster current power supplies for a shorter current pulse with equivalent $i^2r$
heating, or perhaps using several parallel wires running down the decay pipe.  We found that a 2~kA current produces 
20\% more neutrinos in the NuMI low-energy peak than a 1~kA hose.  The 2~kA hose doubles the 'high energy tail'
of the beam, however, so such an ambitious design would be more appropriate for a future, off-axis beamline where
the high energy tail is suppressed.  In any case, a 1~m radius decay pipe with the active focusing of the hose 
at 1~kA produces the same flux as does a beamline with a passive 2~m radius decay volume, and at a reduced cost.

\section{Hose Wire Material Requirements}
\label{requirements}

The choice of wire radius and material must be optimized for several considerations.  First,
larger radius wire tends to reduce neutrino event rates because of pions scattering in the
wire.  Second, larger wire can be better cooled in the near-vacuum of the decay pipe because
both radiative and gas cooling grow with surface area of the wire.  Furthermore, the heating
of the wire is less for larger radius wire because the electrical resistance is reduced and
so is $i^2r$ electrical heating.\footnote{For the current pulse shape chosen, 
the skin depth is of order the wire radius.}  Third, while higher conductivity metals such as
Copper improve electrical resistivity, low $Z$ is desirable to reduce pion scattering and neutrino
event rate losses.  We have chosen Aluminum alloy 1350, whose conductivity is $\approx 80\%$
that of Copper, and a radius of 1.15~mm, as discussed in this section.  The subject of the 
following sections will be to demonstrate that this wire, when anodized to improve its 
emissivity, can survive the environment of the NuMI decay tunnel.

\begin{figure}[t]
\centering
\includegraphics[width=80mm]{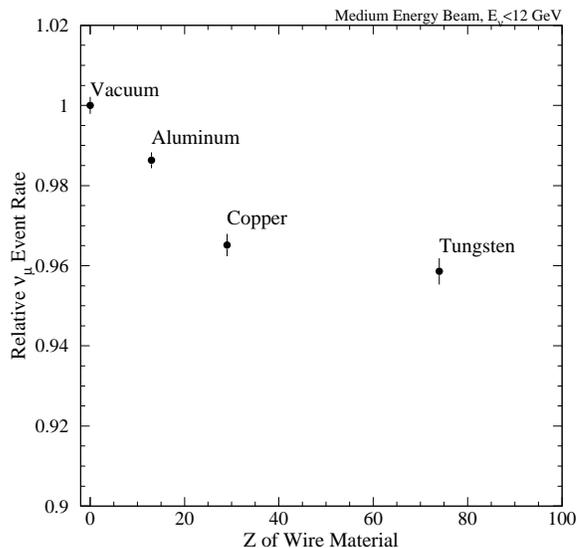}
\caption{Monte Carlo predictions of the 
effect of pion interactions in the wire material on the neutrino event rate.  Shown is the neutrino 
event rate in the MINOS far detector for a hose wire made of 'vacuum' and for several
choices of metals (NuMI medium-energy beam).}
\label{material}
\end{figure}

\begin{figure}[b]
\centering
\includegraphics[width=80mm]{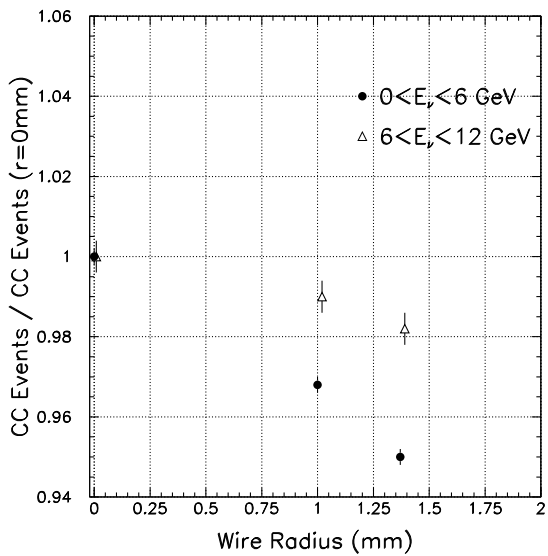}
\caption{Monte Carlo predictions of the effect of the wire radius on the neutrino event rate.  
The simulation assumed an Aluminum hose wire and was simulated for the NuMI medium-energy beam.}
\label{wire-radius}
\end{figure}

To study the effects of wire material, simulations were made using no
wire material, as well as aluminum, copper and tungsten.  The charged-current (CC) event
spectrum at the far detector calculated under these conditions are compared in Figure~\ref{material}. 
The addition of the material results in a 3-6\% decrease in the event rate below 10
GeV. Figure~\ref{material} also shows the change
in the $\nu_{\mu}$ CC event rate at the far detector as a function of the
atomic number of the wire material. In each case the wire radius was
2.8~mm.  As much as a 9\% decrease in the event rate below 12~GeV is
seen for the highest $Z$ material.

The change in event rate is plotted in Figure~\ref{wire-radius} as a
function of wire radius.  The event rates are compared
to simulations using no wire material. All simulations used aluminum
wire. The event rate below 6~GeV is most strongly affected by the wire
radius with roughly a 5\% decrease at a radius of 1.4~mm.  To reduce the loss of neutrino events due to 
scattering in the wire material, we chose a wire radius of 1.15~mm.

\section{Hose Mechanical Design}
\label{mechdesign}

\begin{figure}[t]
\centering
\includegraphics[width=140mm]{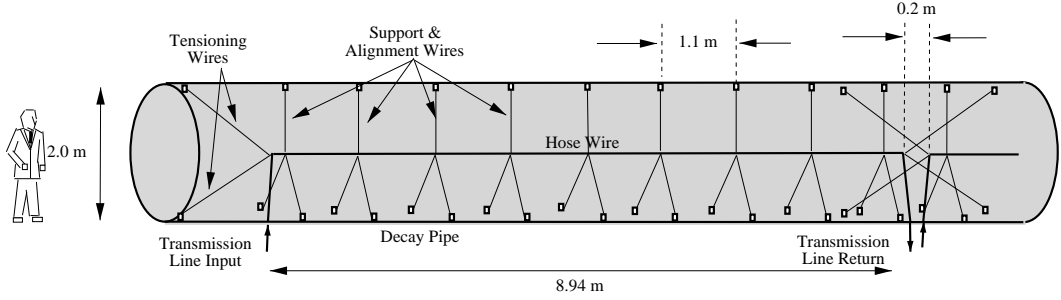}
\caption{Schematic diagram of one hadron hose wire section, indicating support/alignment 
wires and tensioning wires.}
\label{hose-schematic}
\end{figure}

The proposed design is to build 72 sections of wire, each 8.94~m in length and each of which is an 
independent circuit (see Figure~\ref{hose-schematic}).  In this way, the voltage drop across a wire segment is 
reduced and the potential
risk is reduced to the loss of a single segment in the event of a wire failure.  In addition to the 
72 segments, a 2~m long segment at the beginning of the decay pipe acts as a partial shield for the rest 
of the hose from the fraction of the proton beam which does not react in the target.  The 1.15~mm radius Aluminum 1350
alloy wire is anodized with a Type~II aluminum oxide layer 17~$\mu$m in thickness to increase its 
emissivity.\footnote{This is the maximum thickness recommended by Alumat, Inc., the manufacturer.  Studies with
thicker anodization coatings showed that the wire would in fact experience surface cracking in the aluminum when
the wire expands under heat.}
Each hose segment is separated by 20~cm from its neighboring segment.

The hose is supported at the center of the decay pipe by 0.5~mm diameter Invar support wires at intervals 
of 1.1~m along the hose wire's length.  Three Invar wires are plasma-welded to a small Invar loop which 
encircles the hose wire, constraining its position in the vertical and horizontal directions.  Invar is chosen 
because of its low thermal expansion coefficient, so that it maintains hose wire alignment even with 
beam heating.  The hose wire slides
inside the loop with a 0.5~mm clearance, which allows the wire to elongate under heating without misalignment.
The clearance contributes to the alignmnent error of the wire.  

\begin{figure}[t]
\centering
\includegraphics[width=80mm]{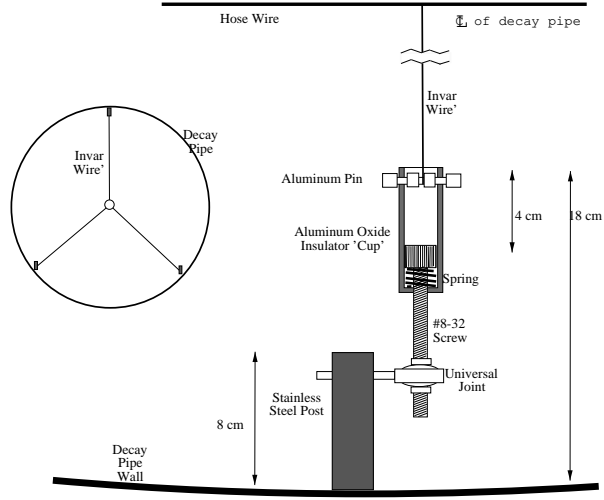}
\caption{Diagram of the invar guide wires which support the hadron hose wire. }
\label{spider}
\end{figure}

The Invar wire is mounted to the 
decay pipe wall with an insulating swivel mount, shown in Figure~\ref{spider}.  A screw in this mount allows
for fine positioning and alignment of the hadron hose wire.  The insulator is chosen to be aluminum oxide
ceramic and is located at the walls to reduce material at the center of the decay volume which can scatter
hadrons.  The choice of materials that can be used as the insulator is limited by radiation levels, ranging from 
10$^{11}$~Rad/year expected at the decay pipe center to 10$^9$~Rad/year expected at the walls of the decay pipe.
A spring inside the lower two mounts tensions the Invar wire at 0.2~N.  The inset to Figure~\ref{spider}
shows the three-fold Invar supports at one point along the hose wire's length.  

At the end of each hose wire segment, a 1.0~cm long, 1.0~cm outer diameter, 2.9~mm inner diameter, hollow 
aluminum cylinder is crimped on to the hose wire.  Invar wires loop around this aluminum cylinder and are 
stretched to springs mounted on the decay pipe wall.  
These springs tension the hose wire to take up the expansion of the wire and reduce wire sag.  
The tensioning springs are mounted to the decay pipe wall using the same assembly as the guide wires.  
Since gravitational sag of the wire is a form of misalignment, the 1.1~m spacing between guide wire supports
suggests that the required tension can be derived from $\delta = 480 L^2 / \sigma$, where $\delta\sim$2~mm 
is the wire sag, $L=1.1$~m, $\sigma$ is the wire tension per unit area in pounds per square inch (PSI), and 
the mass density for aluminum has been assumed.  We find $\sigma~=~290$~PSI, or 2~lbs=0.9~N tension on a 
1.2~mm radius aluminum wire.

\begin{figure}[t]
\centering
\includegraphics[width=75mm]{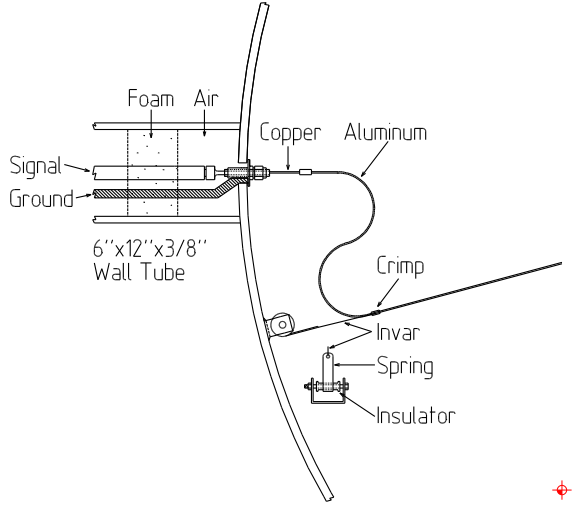}
\caption{Diagram of the electrical feedthrough at the ends of each wire segment.}
\label{feedthrough}
\end{figure}

The current is delivered through the decay pipe wall by a feedthrough constructed of zirconium oxide insulator and Swagelok
compression fittings (see Figure~\ref{feedthrough}).  A pin passing through the hollow 1.9~cm diameter insulator is welded
on the inside to the hose wire segment which is bent out to the decay pipe wall, and on the outside is welded to the 
transmission line.  The insulator is sealed to the wall by a bulkhead Swagelok fitting welded to the wall, and is sealed
to the pin with a 3/4''-to-1/4'' Swagelok reducing union.    A feedthrough is connected to each end of each 9~m
long section.

The effect of misalignment of the Hadron Hose wire was tested in two manners.  First, random
error was studied by introducing random displacements of the wire ends in ($x,y$ and
$z$) in the Monte Carlo simulation.  RMS misalignment of the wire by 3~mm results
in approximately a 2.5\% decrease in the event rate at the far
detector below 6~GeV, and grows to 7\% for misaligments of 5~mm.  We therefore chose
a 2~mm misalignment tolerance.  Second, systematic misalignmnents were studied by misaligning
the entire hadron hose wire with respect to the central beam axis defined by the proton beam, target,
and horns.  In this case, we found that the pion beam orbits follow the misaligned hose direction.  
Thus, collinear offsets of the hose wire result in no change in the neutrino flux, while an angular
misalignments actually result in an off-axis neutrino beam energy spectrum \cite{bnloff} for 
the on-axis detectors.  Other systematic effects of shorter length scale, such
as 'bowed' placement of the hose wire along the decay tunnel, or even gravitational wire sag 
(see Figure~\ref{survey}), tend to result in neutrino flux loss similar to the random misalignments
because of the similar length scale as the random misalignments studied.

The performace of the Hadron Hose has been simulated under various
failure conditions. In the first study, a pessimistic failure rate
of 10\% of the wire segments was assumed and random segments were selected for failure.  
This failure rate causes roughly a 10\% decrease in the near and far detector events
rates.  Simulations were also made assuming failure of the first two
segments. As these segments receive the most beam heating they are
perhaps more likely to fail. Failure of the first two
segments results in roughly a 3\% decrease in the event rate below 8~GeV.  Robustness and the
ability to operate with a few broken segments is important because the decay volume will
become radioactivated, making replacement of a wire impossible.

\section{Full Scale Prototype Section}

We have built a 14~m long prototype of the NuMI decay pipe and instrumented it with a full-length hadron hose
wire segment, as well as two dummy segments at either end.  In addition to practicing installation procedures,
 we investigated whether the wire 
vibrated at all during electrical pulsing, given that the leads of adjacent hose segments could, in principle, 
exert forces on one another of $\mu_0 I^2/2\pi r~\sim~2$~N.  We observed the wire through an optical telescope during
pulsing of the wire.  No effect as large as our sensitivity of 25~$\mu$m was observed.  Furthermore, the wire was
observed to remain centered during various tests to simulate expansion of the wire under heating or creep:  pulling
on one end of the wire segment, the spring tensioning on the opposite end of the segment expanded, while the 
support/alignment wires kept the opposite end centered.  The Invar loop which constrains the hose wire properly 
allowed the wire to slide along the beamline direction, but maintained radial alignment.

\begin{figure}[t]
\centering
\includegraphics[width=130mm]{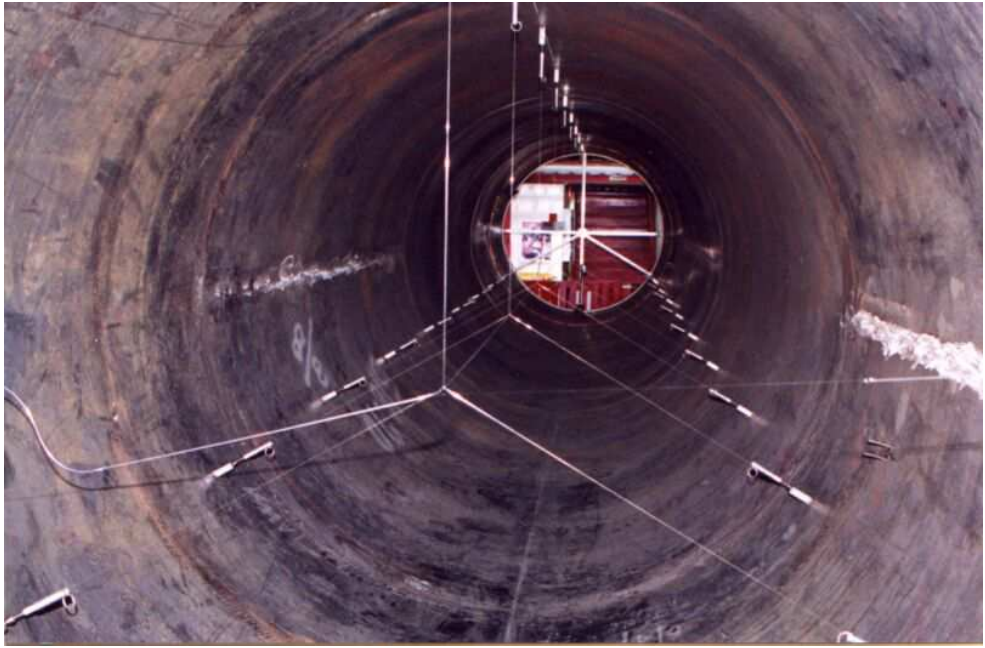}
\caption{Photograph of the full-size NuMI decay pipe segment outfitted with a hadron hose segment.}
\label{proto}
\end{figure}

\begin{figure}[t]
\centering
\includegraphics[width=95mm]{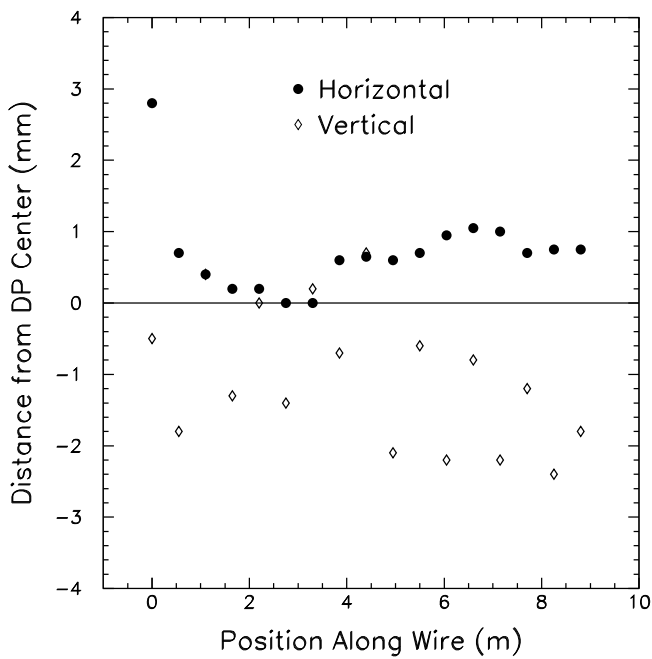}
\caption{Results of the survey of the hose wire in the prototype section after installation.  Measurements
were made at each of the 9 Invar support locations and also halfway in between each pair of supports to
observe wire sag.}
\label{survey}
\end{figure}

The hose wire was aligned by first using a laser tracker to locate all of the support posts which were
welded to the pipe walls.  A 1/4'' hole diameter at the top of each post provided a precision reference point.  The
laser tracker located all the posts within a 10~m span of the pipe, and was then moved down the pipe to survey
an additional, overlapping 10~m span.   These data were then used to define a set of coordinates for the center
of each Invar support with respect to the tooling ball locations in the posts, and the Invar supports 
were set in place to these 
coordinates using the laser tracker.  After sliding the wire through all the Invar support loops and attaching
the tensioning springs, a stick micrometer was used to confirm the relative location of the hose wire to the 
tooling balls mounted on the posts.  The results of the after-installation 
survey are shown in Figure~\ref{survey}.  As can be seen, the horizontal displacements of the wire from the
ideal centerline are within 1~mm, and in the vertical direction the supports are located to within 
1~mm as well.  In the vertical, the observed $\sim~2$~mm displacements of the wire from the centerline which occured
between the invar suuports agree with our expectations for wire sag.

\section{Hose Electrical Design}
\label{elecdesign}

The electrical design of the hadron hose must optimize for three factors.  First, the current pulse of 
1000~A must be long compared to the beam spill length (8.6~$\mu$sec for NuMI) and consistent from wire segment to
wire segment.  Second, the pulse must be fast to reduce $i^2r$ heating of the wire.  Third, the pulse must
be slow enough so that the inductive voltage drop does not cause voltage break down of the wire in vacuum.
This section describes the electrical design that meets the voltage drop and heating requirements described 
in Sections~\ref{breakdown} and \ref{heating}.

\subsection{Circuit Parameters}

The design of the power supply and transmission line to deliver the current pulse
to the hose wires is determined largely by the ciruit parameters for each segment.  
The current pulse will be of order 0.5~msec in duration each 1.8~sec, and runs
down the wire and returns through the iron of the decay pipe wall.

The circuit inductance for the hose is dominated by the vacuum between the wire
and the decay pipe.  The inductance of the wire is
$L_{wire}=\frac{\mu_0}{8\pi}=5\times10^{-2}\mu {\rm henry/m}$.  
The inductance per meter of the vacuum space between the inner
and outer conductors is
$$L_{vac} = {\mu_0 \over 2\pi} \ln({r_{pipe} \over r_{wire}}) 
= 1.34\times 10^{-6}{\rm henry/m} $$
for the radii $r_{wire} = 1.2$~mm and $r_{pipe} =1$~m.
The inductance per meter of the outer conductor, {\it i.e.}: the decay pipe iron,
is small given the skin depth at a frequency of 1500~Hz (the fundamental frequency
of the current pulse for the hose).  The decay pipe iron inductance may be calculated
from\cite{Simonyi}
$$L_{iron} = {1 \over 4 \pi^2 \rho \delta f r_{pipe}} \ 
{\sinh(x)-\sin(x) \over \cosh(x)-\cos(x)}$$,
where $\rho=1.2\times 10^7$~m$\Omega$/m is the resistivity of the iron, 
$\delta = 1/\sqrt{\rho \pi f \mu_{rel} \mu_0}$ is the skin depth of the iron, and 
$x = 2d/\delta$, with $d$ the thickness of the pipe.  Taking a range of values
$\mu_{rel} = $ 500 to 5000, we find $L_{iron}~=~0.03-0.1~\mu$Henry/m.

The electrical resistance of the circuit is dominated by the hose wire.
The electrical resistivity of Aluminum alloy 1350 is $2.96~\mu\Omega$-cm, giving a
series resistance of $R~=~83$~m$\Omega$ for the 894~cm long segments plus 2 one-meter 
leads when at room temperature.  During beam operation the temperature of some segments
will be as much as 150$^\circ$C (see Section~\ref{heating}), increasing the segments' 
resistance to $R~=~120$~m$\Omega$.  The resistance per meter of the iron is 
$$R_{pipe} = {1 \over 2\pi r_{pipe} \delta \rho} \ 
{\sinh(x)+\sin(x) \over \cosh(x)-\cos(x)},$$
which we calculate to be $R_{pipe} \sim 0.1-0.2$~m$\Omega$/m, which is 1000 times smaller than the hose wire.

\begin{figure}[t]
\centering
\includegraphics[width=130mm]{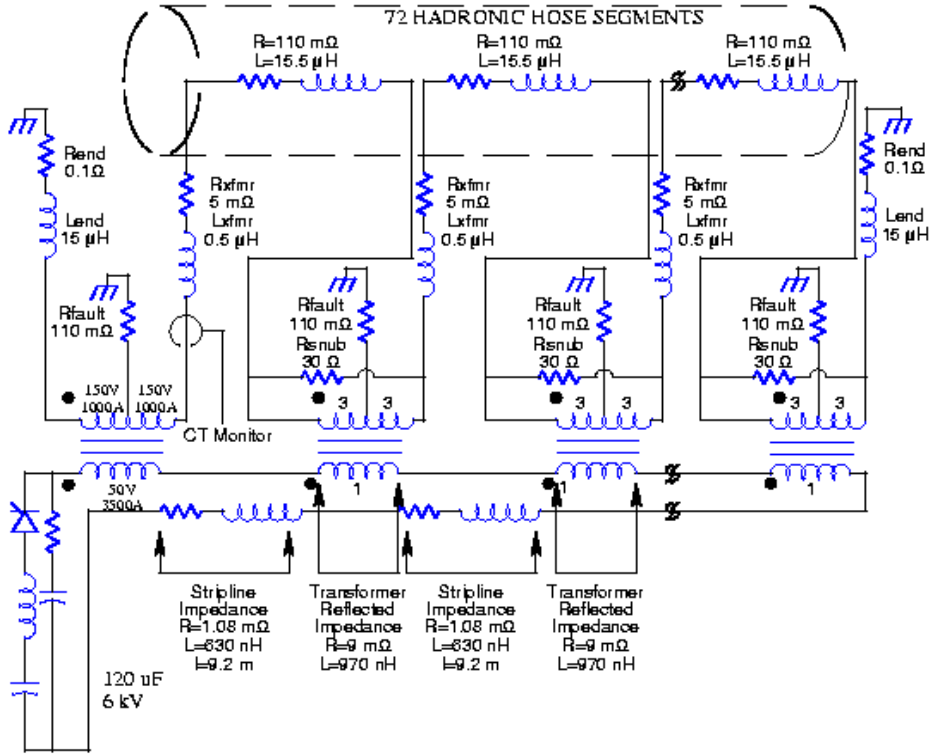}
\caption{The electrical circuit simulated in Spice.\cite{spice}  The 72 
sections are connected to the transmission
line through center-tapped transformers, with the transformers grounded to the decay pipe wall.  A dummy 
impedence terminates the first and last transformer.}
\label{circuit}
\end{figure}

Because the decay volume is embedded in 2.5-3.0~m of concrete, it is necessary to have long leads from the
power transmission line to the hadron hose feedthroughs.  
These leads consist of parallel flat copper sheets, and contribute an additional 2.5~$\mu$Henry 
inductance and 5~m$\Omega$ resistance in series with the hose segments.

\subsection{Hose Circuit}

Each hose segment is connected in parallel to a transmission line that runs down the side
passageway of the decay tunnel.
The transmission line is energized to 5000~V.  The hose wire segments are connected through 73
3:1 transformers, with the upstream
end of a hose segment center-tapped to the same transformer as the downstream end of the
preceding hose segment.  Dummy impedance loads are connected to the first and last transformers (see
Figure~\ref{circuit}).  Current tranformers around the hose wire leads on the hose side of the transformer
are used to monitor the current through each hose section and check for wire shorts or breaks.

The center-tapping of the transformers accomplishes three objectives.  First, the voltage of any wire
relative to the grounded vacuum decay pipe is cut by a factor of two, reducing the risk of voltage breakdown.
Second, all the hose segments are effectively in series and receive the
same currents to within 1~A according to simulations.  Third, if a hose segment should fail
and sever during operation, all the current is bypassed through the transmission line to the next hose segment.

\subsection{Transmission Line}

\begin{figure}[t]
\centering
\includegraphics[width=140mm]{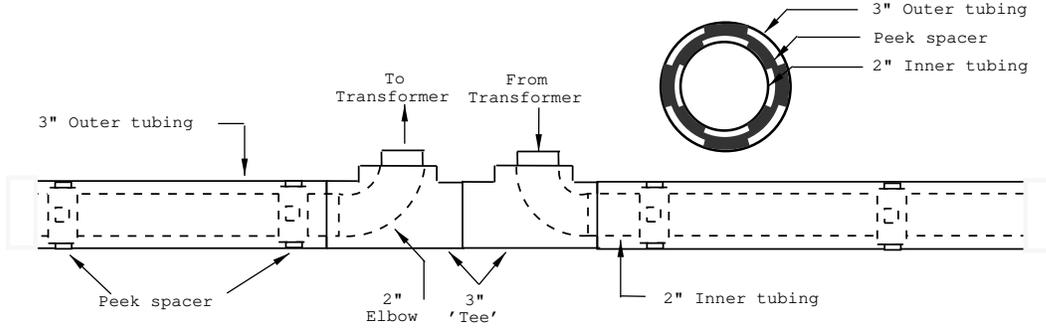}
\caption{Schematic drawing of the coaxial transmission line in the decay tunnel.  Concentric 3'' and 2'' outer
diameter Copper tubing are spaced from one another by Peek insulating rings.  Connections between the center 
transmission line conductor and the transformers are accomplished by welding 3'' Copper tubing ``Tee's'' to the
end of each outer conductor segment, and a 2'' ``elbow'' to the end of each inner conductor segment.}.  
\label{transline}
\end{figure}

Because of the ambient conditions of the decay tunnel passage, we designed a coaxial transmission
line made of 2 inch and 3 inch outer diameter concentric tubes of commercially available copper tubing.  The center
hot conductor is spaced from the outer return by circular standoffs spaced every meter along the transmission
line.  The standoffs are made of Peek plastic.  With this design, no protection from ambient water or humidity is
necessary.  Measurements made on several prototype segments indicate that commercially available
copper tubing (schedule 'K') has a resistance of 1.6~m$\Omega$ per 9~m length and the concentric tubes have 
an inductance 900~$\mu$H.  We also found that poor-quality weld joints can dominate
the resistance of each 9~m transmission line segment.
A schematic drawing of the transmission line design is shown in Figure~\ref{transline}.

\subsection{Circuit Simulation}

We have simulated the hadron hose circuit using Spice, including the inductance 
and resistance of the hose wire
segments, penetration leads through the concrete, and transmission line.  Furthermore, the Spice model 
included variations in the series resistance between hose wire segments due to the different temperatures
of the hose wires at different positions along the beamline.  A schematic of the simulated circuit is 
shown in Figure~\ref{circuit}.

\begin{figure}[t]
\centering
\includegraphics[width=105mm]{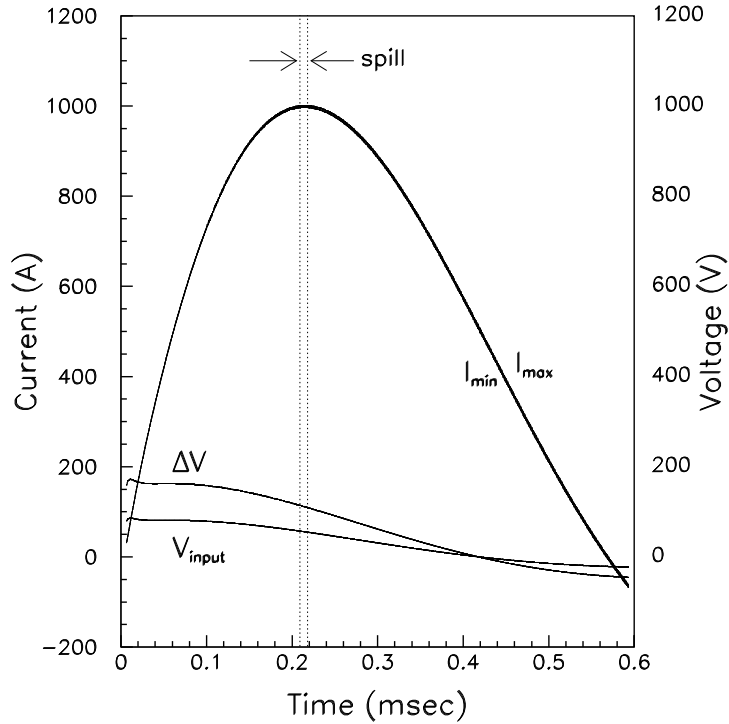}
\caption{Spice simulation of the current and voltage pulse on a hadron hose wire segment.  Indicated
are the input voltage $V_{\mbox{input}}$ on each wire segment, the voltage difference $\Delta V$ between upstream
and downstream end of the wire, and the currents for the hottest and the coolest of the wire segments
in the beamline.}
\label{spice}
\end{figure}

The results of the simulation,shown in Figure~\ref{spice}, have 3 important features.  First, the variation 
of the delivered current during the beam spill is 1~Amp, or 0.1~\%.  Second, the voltage
of the input ($V_{\mbox{input}}$ in Figure~\ref{spice}) is no more than 150~V anywhere, and the voltage difference
between input and output ends ($\Delta V$ in Figure~\ref{spice}) is no more than 250~V.  This means that 
the most challenging location for voltage breakdown is restricted to the 20~cm gap between hose wire segments.
Third, while the temperatures of the hose segments in the decay volume vary from 80$^\circ$~C at the downstream
end to 150$^\circ$~C near the upstream end of the decay tunnel (see Section~\ref{heating}), it is seen that the
segments' currents vary by less than 1~Amp.

\section{Voltage Breakdown}
\label{breakdown}

\begin{figure}[t]
\centering
\includegraphics[width=90mm]{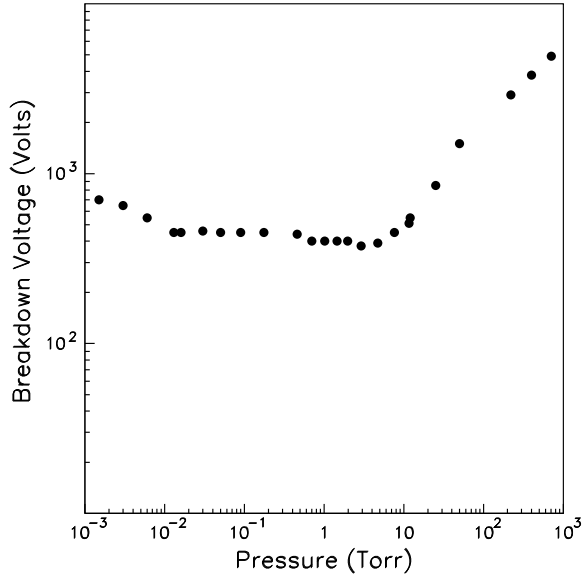}
\caption{Breakdown voltage of a hose wire segment inside a 6'' diameter vacuum chamber as a function
of the vacuum chamber pressure.}
\label{breakdownplot}
\end{figure}

The hose wire will have a voltage drop across each segment of 250~Volts.  This voltage inside the 
0.1~-~1.0~Torr vacuum of the NuMI decay pipe poses two potential problems.  First, if the potential
drop is too large then electrical breakdown could occur between the hose segments, which are separated 
by $\sim$~20~cm.  As shown in Figure~\ref{breakdownplot}, the Paschen curve \cite{paschen} reaches a 
minimum of 400~Volts in the pressure range of 0.1~-~1.0~Torr relevant for the NuMI decay volume.  Limiting
the voltage drops inside the decay volume to 250~V provides a margin of safety against voltage breakdown.

The second problem is the ionization of gas in the decay pipe
from charged particles in the beam.    
This residual gas ionization could result in an electrical breakdown or in a current deposited on the wire which 
modifies the expected current pulse.  Assuming the particle flux through the decay pipe is
$6~\times~10^{13}$~particles/spill, an energy deposition of $dE/dx_{min} = 1.82$~MeV/(g/cm$^2$) in air, 
and scaling for the 0.17-1.7~mg/l density of air at the 0.1-1.0~Torr decay pipe pressure, 
the potential current collected on a 9~m hose wire segment is in the range between 9 to 90~Amp, 
ignoring effects of ion mobilities, recombination, etc.

We measured the ion current collected on a hose wire segment from ionized residual gas in a beam test
at the Fermilab Booster Accelerator.  
An 8~GeV proton beam was passed through a vacuum chamber with 75~$\mu$m thick Ti entrance and exit 
windows.  Inside the vacuum chamber, segments of hose wire segments 
were placed at spacings of 20~cm and 10~cm apart from one another, representing the 
distance between hose segments in NuMI before and after 10~years' worth of creep.  The beam
passed directly between two wires, so that ionization between the wires would drift to the two
electrodes, one at voltage and the other at ground.  
The data presented here were taken at 4.4$\times 10^{12}$
particles per 1.56~$\mu$sec spill, with $\sim 10$\% variation between spills.  
The beam spot size was 
approximately 1~cm RMS, as measured by profile chambers which were retracted for most of the
run.  The ionization current was measured using a Pearson model 4100 current readout toroid around the 
high voltage leads to the chamber and read out in a digitizing scope.

\begin{figure}[t]
\centering
\includegraphics[width=65mm]{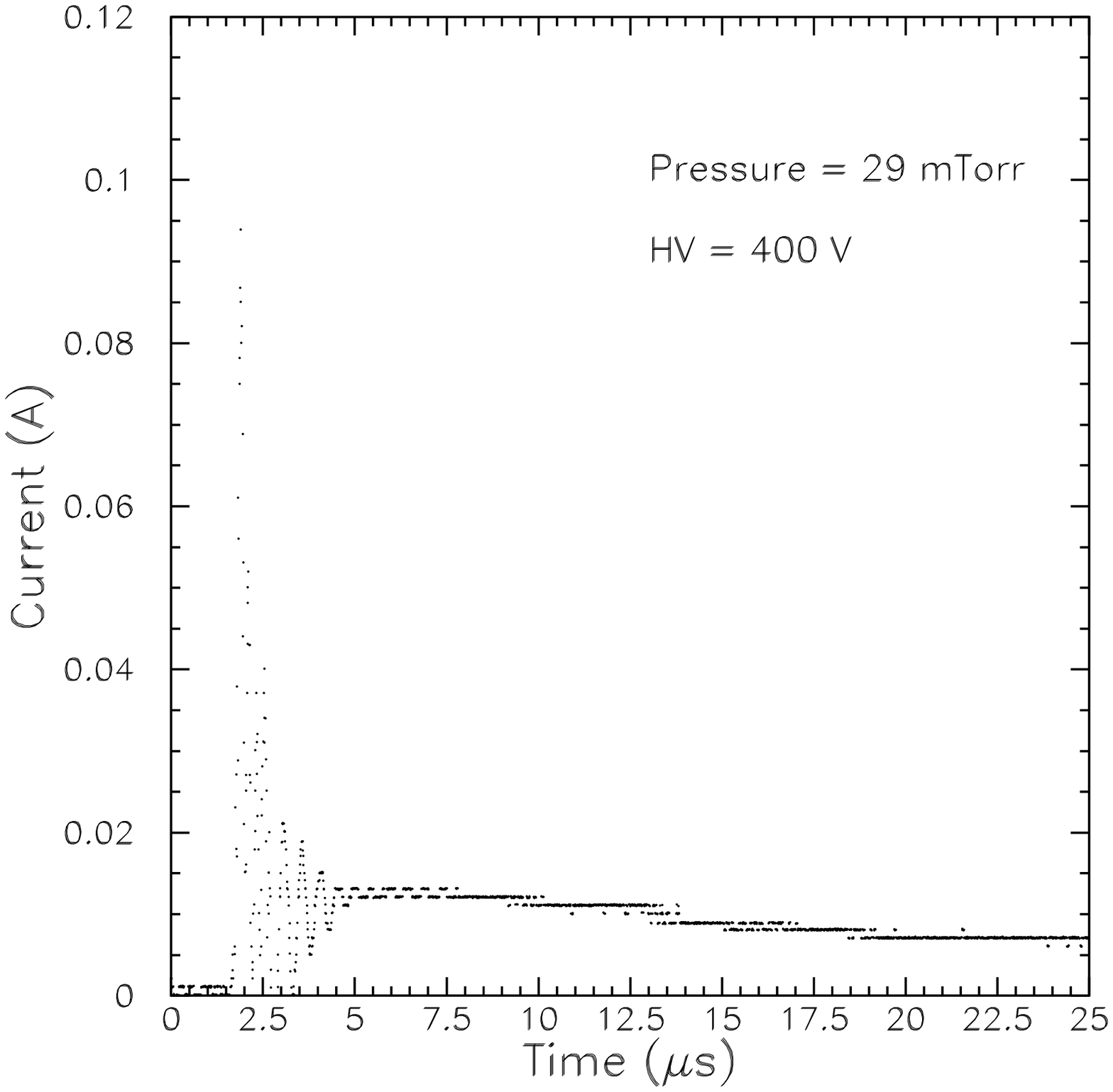}
\includegraphics[width=65mm]{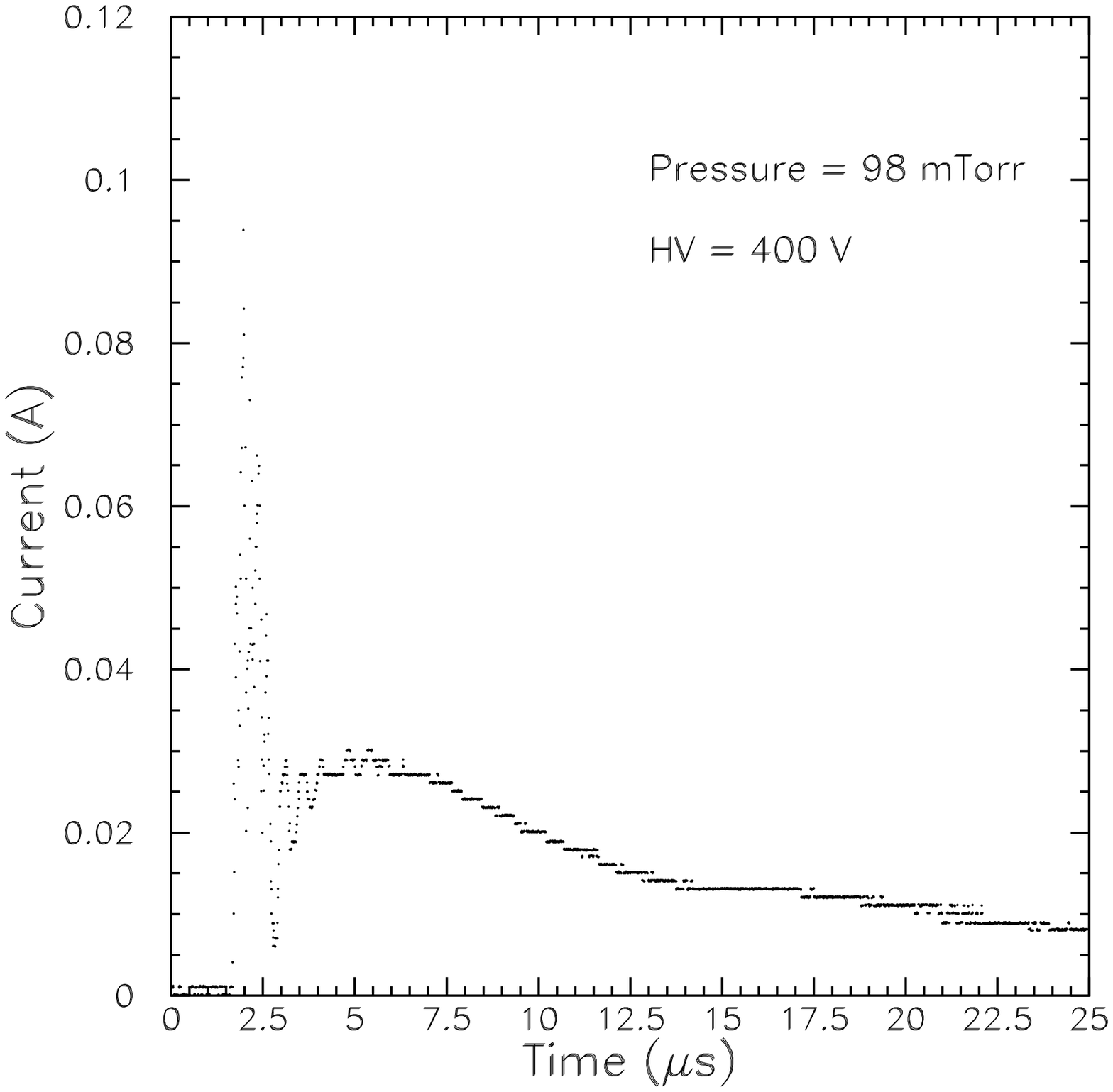}
\caption{Oscilloscope traces of two beam spills passing through the hose bell jar.  The potential
difference between the wires and the pressure inside the bell jar are indicated.}
\label{rdfspills}
\end{figure}

\begin{figure}[b]
\centering
\includegraphics[width=65mm]{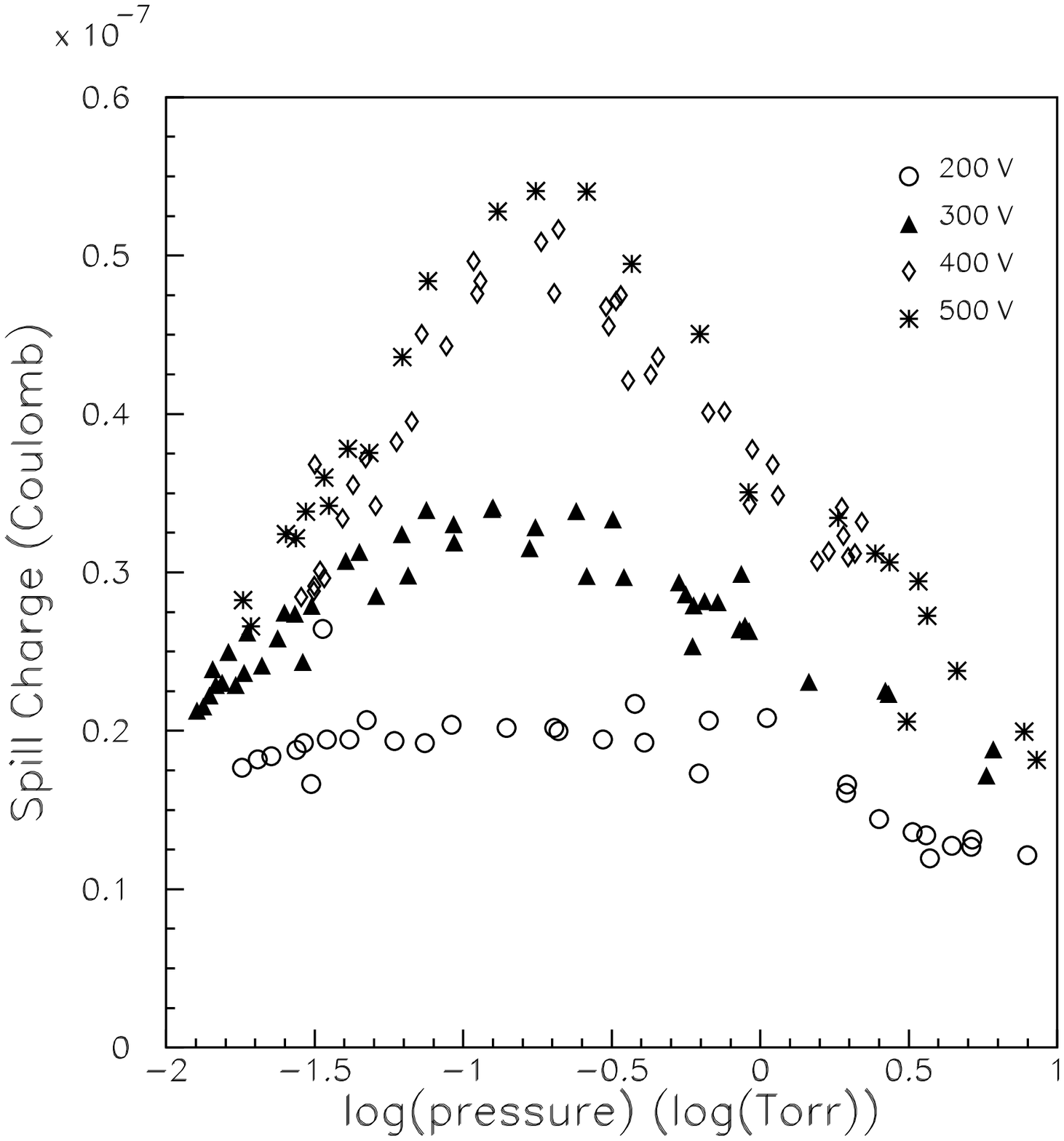}
\includegraphics[width=65mm]{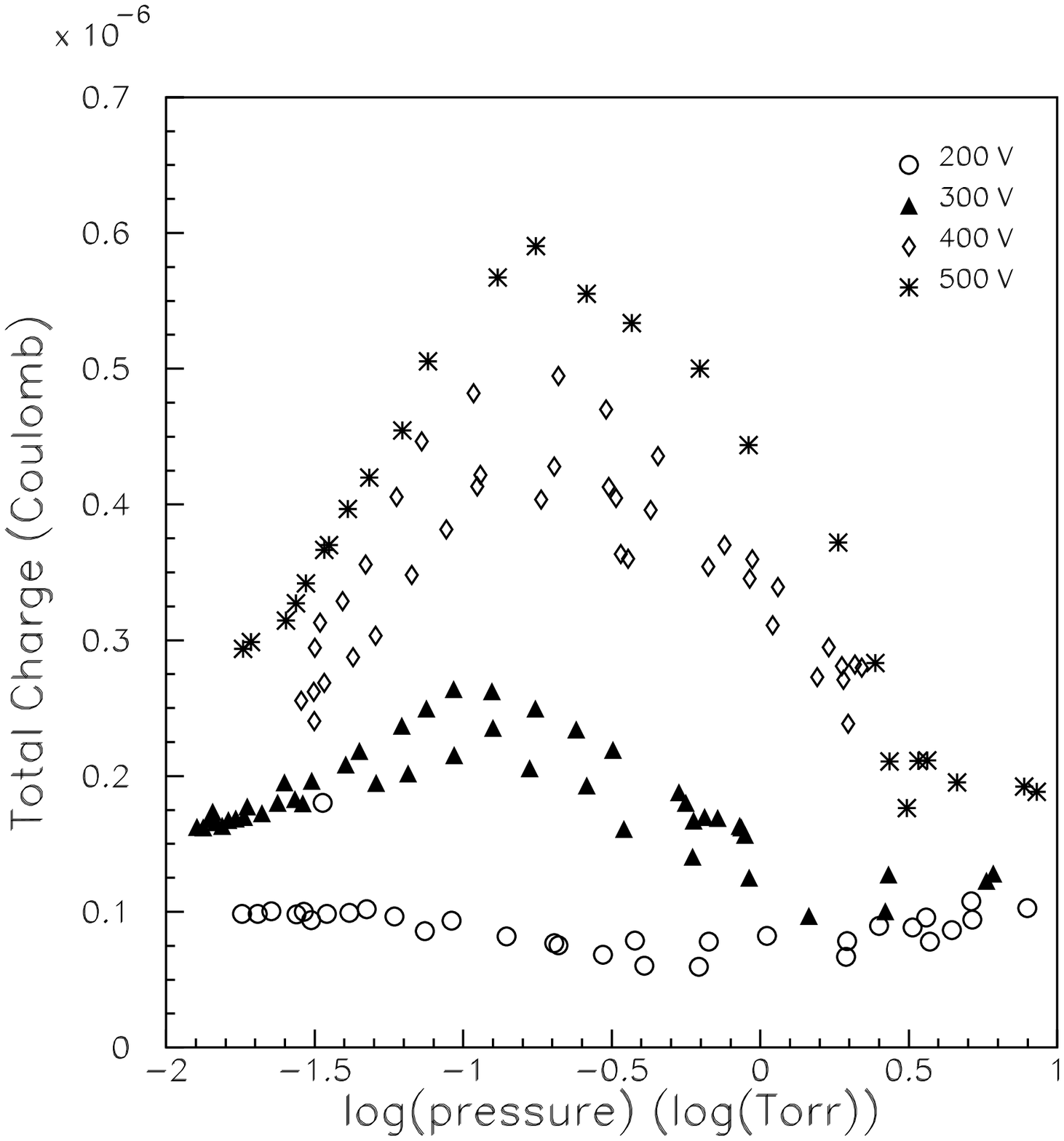}
\caption{Total charge collected on the hose wire during the beam spill (left plot) and 
integrating for 25$\mu$sec after the beam spill (right plot) at several wire voltages and
chamber pressures.}
\label{rdfdata}
\end{figure}

Figure~\ref{rdfspills} shows two typical events at different vacuum chamber pressures 
from the beam test taken at 400~V potential 
difference between the 10~cm-separated wires.  The 1.56~$\mu$sec beam spill occurs between 1.75-3.25~$\mu$sec.
The sign of the current is positive for electron flow back to the power supply from the hose wire, consistent 
with what would be expected for electron current flow from the gas to a positive voltage wire and ion current 
flowing to the grounded wire.  In Figure~\ref{rdfdata}, the total charge is integrated over the 1.56~$\mu$sec 
spill time and over the 25~$\mu$sec oscilloscope sweep time for each event.  Data at several potential
differences and pressures was recorded.  It is perhaps interesting to note that the 200~V data shows no
particular accentuated behaviour at moderate pressures, while an increasingly large charge is collected
at moderate pressure for larger voltages.  This effect perhaps indicates the onset of electrical  
breakdown consistent with the Paschen curve.  

The data from Figure~\ref{rdfdata} may also be compared to the naive calculation of expected ionization 
current in air.  Using the same numbers as above and the 66~cm path length of the beam through the
vacuum chamber, we would expect 0.42~$\mu$Coulomb of charge released in the residual gas at 
100~mTorr.  This estimate agrees roughly with our data for 500~V, but is invalid at higher
pressures or lower voltages, indicating that recombination effects are important.  Furthermore,
only 10\% of the collected charge arrives during the beam spill time (evidently the
drift time is long).  Extrapolating this data to the NuMI beam conditions, the 
current deposited on the hose wire during the NuMI beam spill will be 0.02-0.2~Amps, which is well below
the pulsed current of 1000~Amps.

\section{Wire Creep}
\label{creep}

At temperatures of order 1/3 the melting point, many materials under strain experience plastic flow, or 'creep'.
The melting point of pure aluminum is 640$^\circ$C, suggesting that creep is a concern for the hose wire at
temperatures near 200$^\circ$C.
The effect of creep can be detrimental to the hadron hose for two reasons:  (1) it can lead to the failure of 
a hose wire segment which cannot be replaced after the beam has run;  (2) the hose wire segments could break
down electrically to adjacent segments if creep causes the wires to stretch too close to neighboring wires.
We have performed two measurements of creep of anodized Aluminum alloy 1350 wires and determined a maximum
desirable operating temperature of 150$^\circ$C.

\begin{figure}[t]
\centering
\includegraphics[width=100mm]{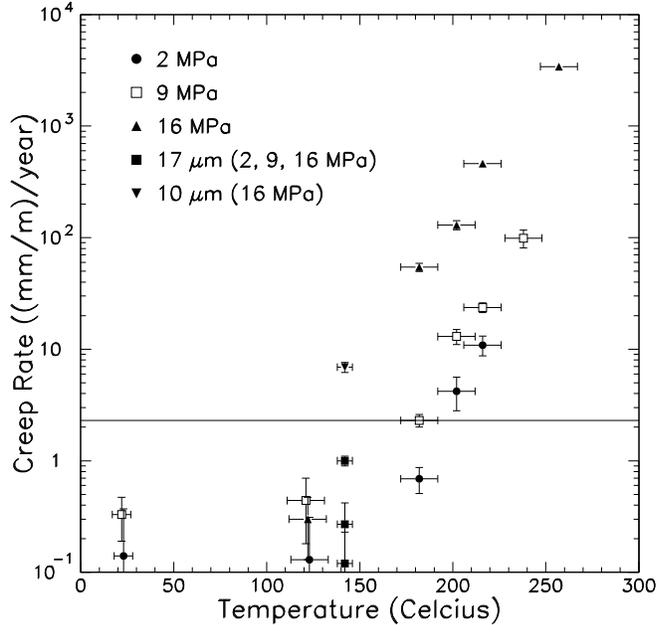}
\caption{Summary of creep data for Aluminum 1350.  The upper 3 plot symbols refer to wires with 10~$\mu$m anodization
thickness and were taken with the 1~m ovens over a period of a year.  The lower two refer to data from the 
13~m oven with wires of  with 10~$\mu$m or 17~$\mu$m thick anodization layer.
The horizontal line at 2.5 mm/m/year indicates the requirement that the hose sections not be permitted to
be closer than 10~cm after 10 years of NuMI operation.}
\label{creepsummary}
\end{figure}

The two measurements of Al creep rates were performed by heating aluminum wires under tension.  The wires were
heated by placing them inside concentric steel tubes which were separated by high temperature plastic.  The 
inner tube was wrapped in heating tape, and the outer wrapped in fiberglass insulation.
Thermocouples monitored the interior chambers.  Wires suspended inside the tube were fixed at one end and at
the other protruded outside the oven to a brass weight which kept the wire under tension.  Dial indicators 
tracked the location of the weights over time.  The first setup consisted of 1~m long ovens which were 
operated for 1~year.  The second setup was a 13~m oven containing 30~wires and operated for one month.
In the small ovens, anodized wire with 10~$\mu$m thick Aluminum oxide coating was tested.  
In the large oven, wires with both 10~$\mu$m and 17~$\mu$m thick anodization layers were tested.
The large oven was used as well to straighten segments of hose wire for installation in the NuMI beamline,
by annealing out the coil from the spool.

Figures~\ref{creepsummary} shows the data collected from all the 1~m tubes.  In the small oven, 
large creep rates were measurable for
temperatures greater than 180$^\circ$C, the data below this temperature were consistent with zero creep.
Better sensitivity was achieved with the large oven, in which a
creep rate of 1.8$\times10^{-4}$/year was observed with the 17-$\mu$m thick coatings at 290PSI~=~4~MPa tension
at 145$^\circ$C.  The thicker anodization appeared to have a factor of 5 or more smaller creep rate than the
10$\mu$m anodization.

\section{Hose Thermal Measurements}
\label{cooling}

Heat dissipates from the hose wire mainly by blackbody radiation and by conduction through the residual 
decay pipe atmosphere to the outer decay pipe wall.  We performed measurements of heat dissipation from the hose 
wire via both of these mechanisms.  These measurements serve as inputs to a detailed thermal model of the hose
wire during beam operation.

\begin{figure}[t]
\centering
\includegraphics[width=80mm]{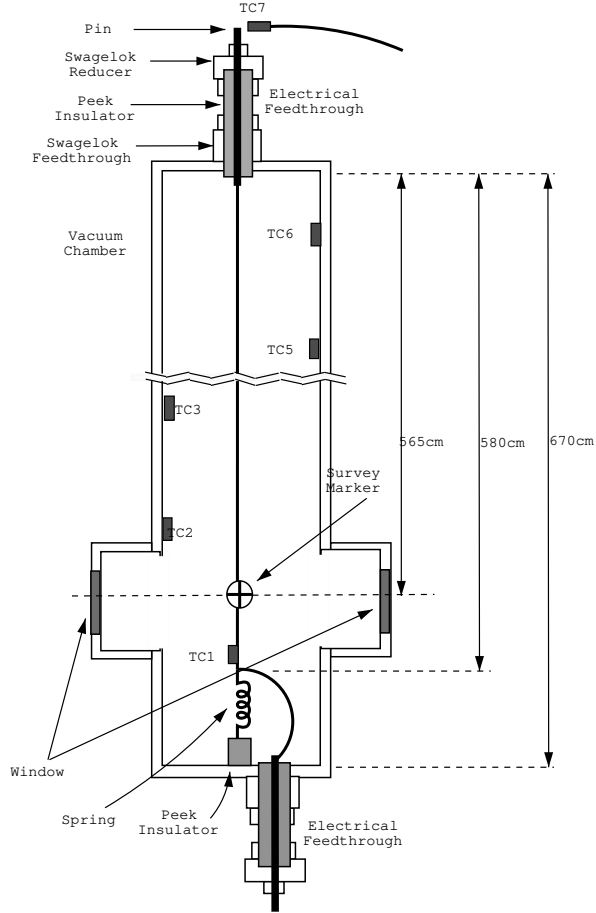}
\caption{Schematic diagram of the vacuum chamber used to measure the emissivity of coated wires and study the 
effect of gas cooling.  A wire is suspended under tension and heated under a known current.  
The heated wire's elongation is measured via a survey marker on the wire. TC's 1-8 are thermocouples.}
\label{emisschamber}
\end{figure}

The thermal measurements of hose wire heat dissipation are performed using 
a vacuum chamber (see Figure~\ref{emisschamber}) in which a known current is passed through a 
hose wire segment suspended under tension at its central axis.  The inside walls of the vacuum chamber are lined with 
black plastic.  The controlled $i^2r$ power input is balanced by blackbody 
radiation to the walls, by conduction through the wire to its ends, and
(possibly) by conduction through the chamber gas to the walls.  The wire comes to
equilibrium at a temperature which is monitored by its elongation.  Viewports to the vacuum chamber
allow monitoring of the elongation through an optical telescope.

To interpret the experimental data on wire elongation, a thermal model of the vacuum chamber and the 
wire is developed.  The temperature rise of the wire in a small time interval $\Delta t$ is
\begin{equation}
\Delta T = \frac{1}{m C_p}(P_{in} - P_{out}^{rad} - P_{out}^{gas} - P_{out}^{cond}) \Delta t,
\label{balance}
\end{equation}
where $P_{out}^{rad}$ is the blackbody radiated power, $P_{out}^{gas}$ is the power conducted through the 
chamber gas (=0 at sufficiently low pressures), $P_{out}^{cond}$ is power conducted through the wire
material to the ends, $m$ is the wire mass, and $C_p$ is its heat capacity.

The input power to the wire $P_{in} = i^2 \rho l /A_{wire}$, where $i$ is the applied current,
$l=580$~cm is the length and $A_{wire}=0.042$~cm$^2$ is the wire cross sectional area, and 
$\rho=2.96~\mu\Omega$-cm is our measured value of the resistivity of Al1350 at 20$^\circ$C.  
The resistivity slope is $0.010~\mu\Omega$-cm/$^\circ$C.  We used our measured value for the coefficient 
of thermal expansion of the anodized wire, $2.25\times10^{-5}$.\footnote{Our measurement for the CTE is lower than the value of $2.4 \times 10^{-5}$ for Aluminum 
1350 \cite{alum_cte}, presumably because of the 17~$\mu$m thick anodization layer on the wire.}

\begin{figure}[t]
\centering
\includegraphics[width=65mm]{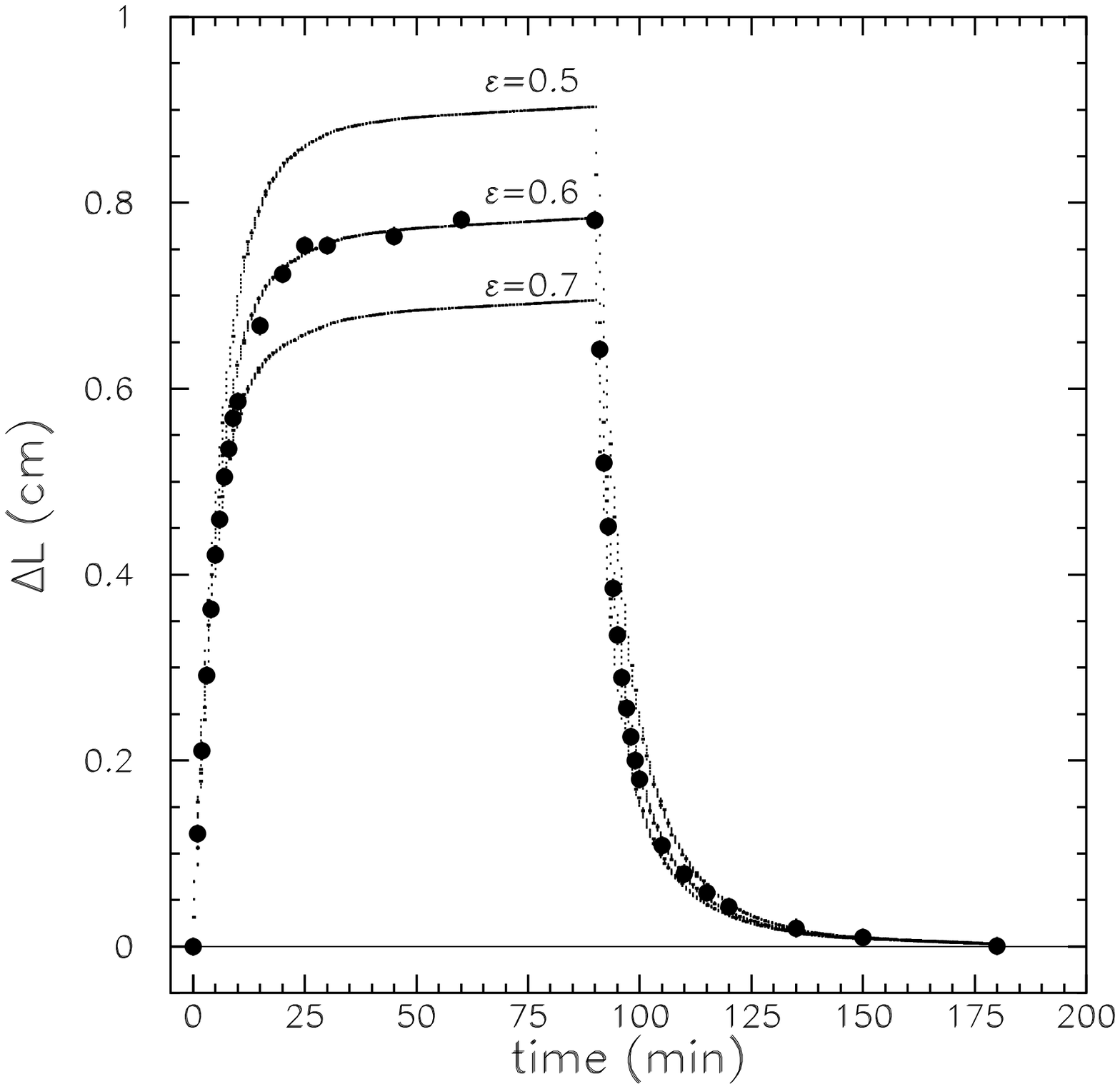}
\includegraphics[width=65mm]{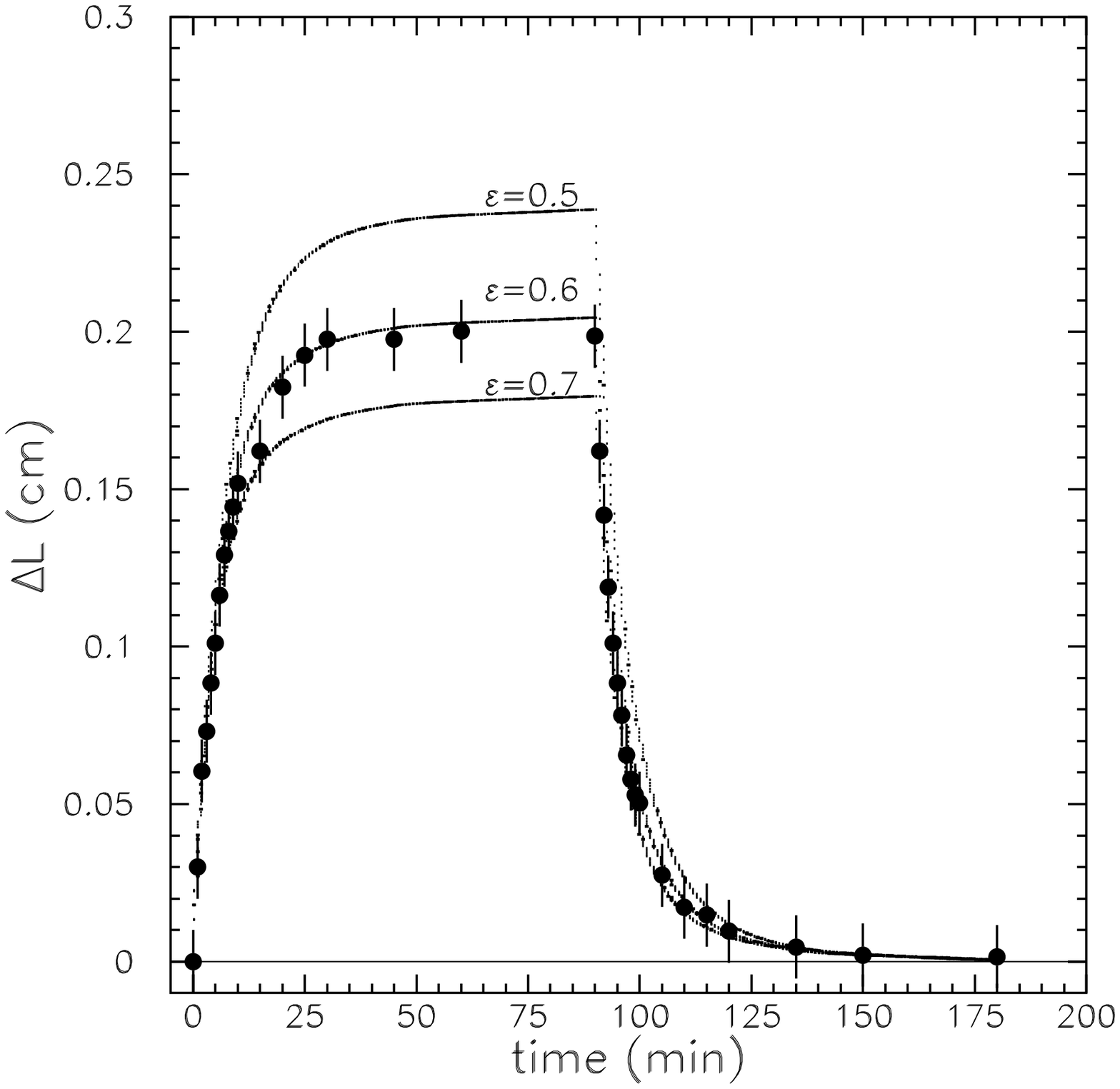}
\caption{Observed elongation of the 580~cm long wire vs time with $4.5 \times 10^{-6}$~Torr 
pressure in the vacuum chamber.  The current was switched on at $t=0$~min and switched off at 
$t=90$~min.  The curves are from the thermal model.  The current on the wire is 15~Amps (left)
and 7.5~Amps (right).}
\label{elongation20}
\end{figure}

The blackbody power radiated is described by
\begin{equation}
P_{out}^{rad} = \sigma_B A_{wire} F_{12} (T_{wire}^4 - T_{wall}^4),
\end{equation}
where $\sigma_B$ is the Stefan-Boltzmann constant, $A_{wire} = 2 \pi r \Delta l$ is the surface area,
$T_{wire}$ ($T_{wall}$) is the temperature of the wire (vacuum chamber wall).  The factor $F_{12}$ is given by 
\begin{equation}
\frac{1}{F_{12}} = \frac{1}{\varepsilon_{wire}} + \frac{A_{wire}}{A_{wall}}(\frac{1}{\varepsilon_{wall}} -1),
\end{equation}
where $\varepsilon_{wire}$ is the wire emissivity to be measured in this study, $A_{wall}$ is the 
surface area of the vacuum chamber walls, and $\varepsilon_{wall}$ is the wall emissivity.
Because $A_{wire}/A_{wall} \sim 0.01$, we placed a black polyethelene liner
inside the vacuum chamber to keep $F_{12} \approx \varepsilon_{wire}$ to within $\sim~0.5$\%.
The plastic liner caused imperfect thermal contact of the walls with the room, so that $T_{wall}$ was not
simply room temperature.  $T_{wall}$ was measured using thermocouples placed inside the vacuum chamber and
on the wire ends (see Figure~\ref{emisschamber}).

Four measurements were made with $i~=~7.5, 10, 15,$ and 20~Amps, yielding emissivity values of $\varepsilon~=~$0.61, 
0.59, 0.60, and 0.61, respectively.  The results of the measurements with $i~=~15$ and 7.5~Amps are
shown in Figure~\ref{elongation20}.  The systematic uncertainties on this measurement
include a $\pm0.01$ uncertainty from the 1\% measurement uncertainty in the coefficient 
of thermal expansion and a $\pm0.01$ uncertainty from the 1\% measurement uncertainty in the electrical
resistivity.  The simulation indicates that the thermal conductivity and heat capacity uncertainties
lead to negligible uncertainty on $\varepsilon_{wire}$.  We
quote $\varepsilon_{wire}~=~0.61~\pm~0.03$ for the wire tested with 17$\mu$m anodization layer.

\begin{figure}[t]
\centering
\includegraphics[width=90mm]{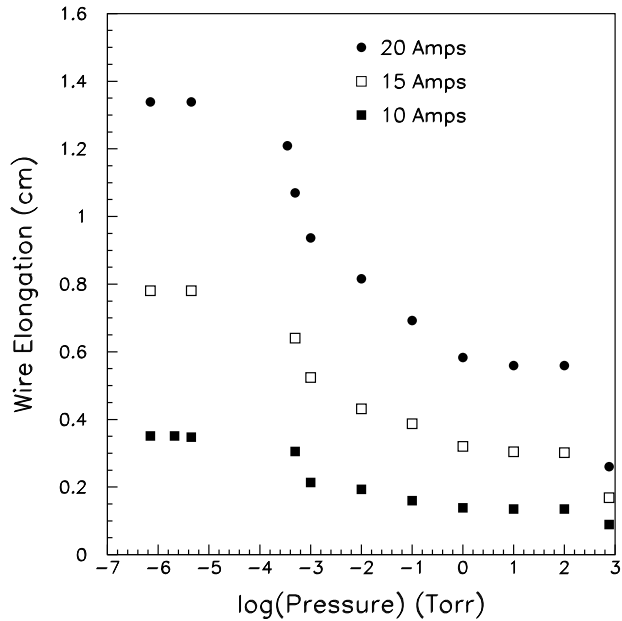}
\caption{Observed elongation of the 580~cm long wire for several vacuum chamber pressures and three different
wire currents.}
\label{l_vs_p}
\end{figure}

\begin{figure}[t]
\centering
\includegraphics[width=90mm]{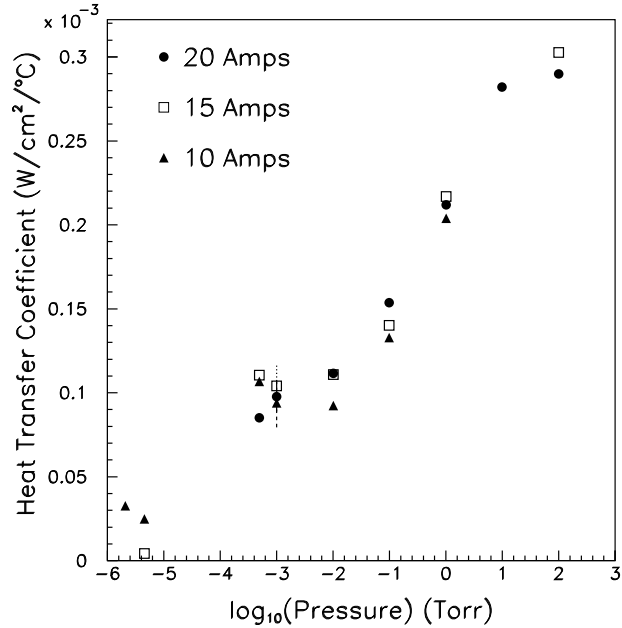}
\caption{Derived values for the gas heat conduction coefficient versus pressure from the data of
Figure~\ref{l_vs_p}.}
\label{htrans}
\end{figure}

The same chamber was used to measure the effect of gas cooling.
Figure~\ref{l_vs_p} shows several measurements of wire elongation at different currents and vacuum
chamber pressures.  Apparently gas cooling is important above 
$10^{-5}$~Torr, and the term
$P_{out}^{gas}$ in Equation~\ref{balance} must be included.  We assume a form of 

\begin{equation}
P_{out}^{gas} = \frac{2\pi k}{ln(r_{wall}/r_{wire})} \Delta l (T_{wire} - T_{wall}),
\label{gas}
\end{equation}

where $\Delta l$ is the length of an element of the wire into which the wire is divided, and
$k$ is the heat conduction coefficient of air, which is 0.024~W/cm/$^o$C at
atmospheric pressure.  We use the data in Figure~\ref{l_vs_p} 
to derive $k$ as a function of pressure, setting $\varepsilon_{wire}~=~0.61$.
This is shown in Figure~\ref{htrans}.  The rise of our apparent value for $k$ above 0.0239~W/cm/$^\circ$C
at 10~Torr may indicate that convection is additionally important as the pressure increases.
These values for $k$ are used to simulate the cooling of the hose in the NuMI decay pipe.

\section{Hose Wire Thermal Modelling}
\label{heating}

During operation of the beam, the energy deposited in the Hadron Hose wire comes from $i^2r$ heating 
and from particle interactions in the hose wire.  The hose wire dissipates heat primarily
through  blackbody radiation and cooling from the residual gas in the decay volume.  Measurements of 
these two effects were presented in the previous section.  Cooling via conduction to the wire ends
also occurs, but to only small effect.
In this section we calculate the final equilibrium temperature of the wire during beam 
operation.  Assuming a 150$^\circ$C upper limit for the operating temperature temperature of the wire to
limit the effects of long-term creep, we calculate the maximum current pulse length acceptable to be 0.6~ms.

\begin{figure}[t]
\centering
\includegraphics[width=120mm]{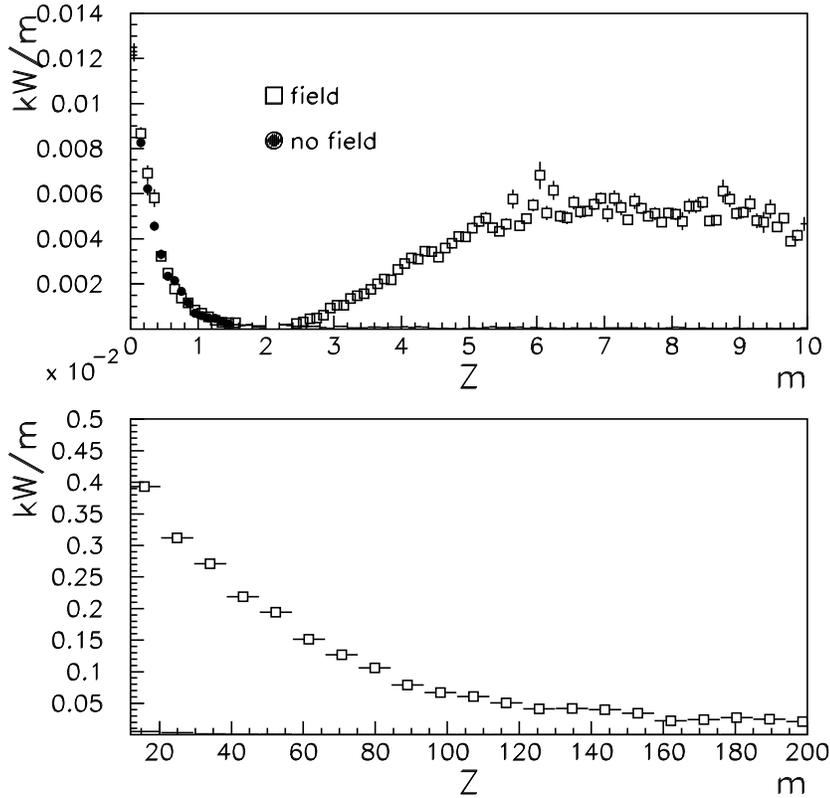}
\caption{The energy deposited along the hose wire due to ionizing particles in the beam as 
calculated using the MARS Monte Carlo \cite{MARS}.  The calculation is performed both for the case of 
no current in the hose and with the magnetic field of the hose turned on.  The peak at zero is due to 
protons striking a 2~m unpulsed section of wire placed in front of the pulsed sections..}
\label{beamheating}
\end{figure}

In one pulse, the current deposits a heat load of 
$\Delta Q = \int i(t)^2 r dt$ delivered to the wire, which grows linearly with the pulse duration.
For an RMS current duration of 310 (620)~$\mu$sec, the simulation of the hadron hose circuit indicates that 
$\int i(t)^2 dt = 215$~(430)~Amp$^2$-sec.  At room temperature, this value for the energy deposited 
would correspond to 18~(36)~J
deposited in one 9~m long hose segment per beam spill, corresponding to 1~(2)~Watt/meter.

\begin{figure}[t]
\centering
\includegraphics[width=65mm]{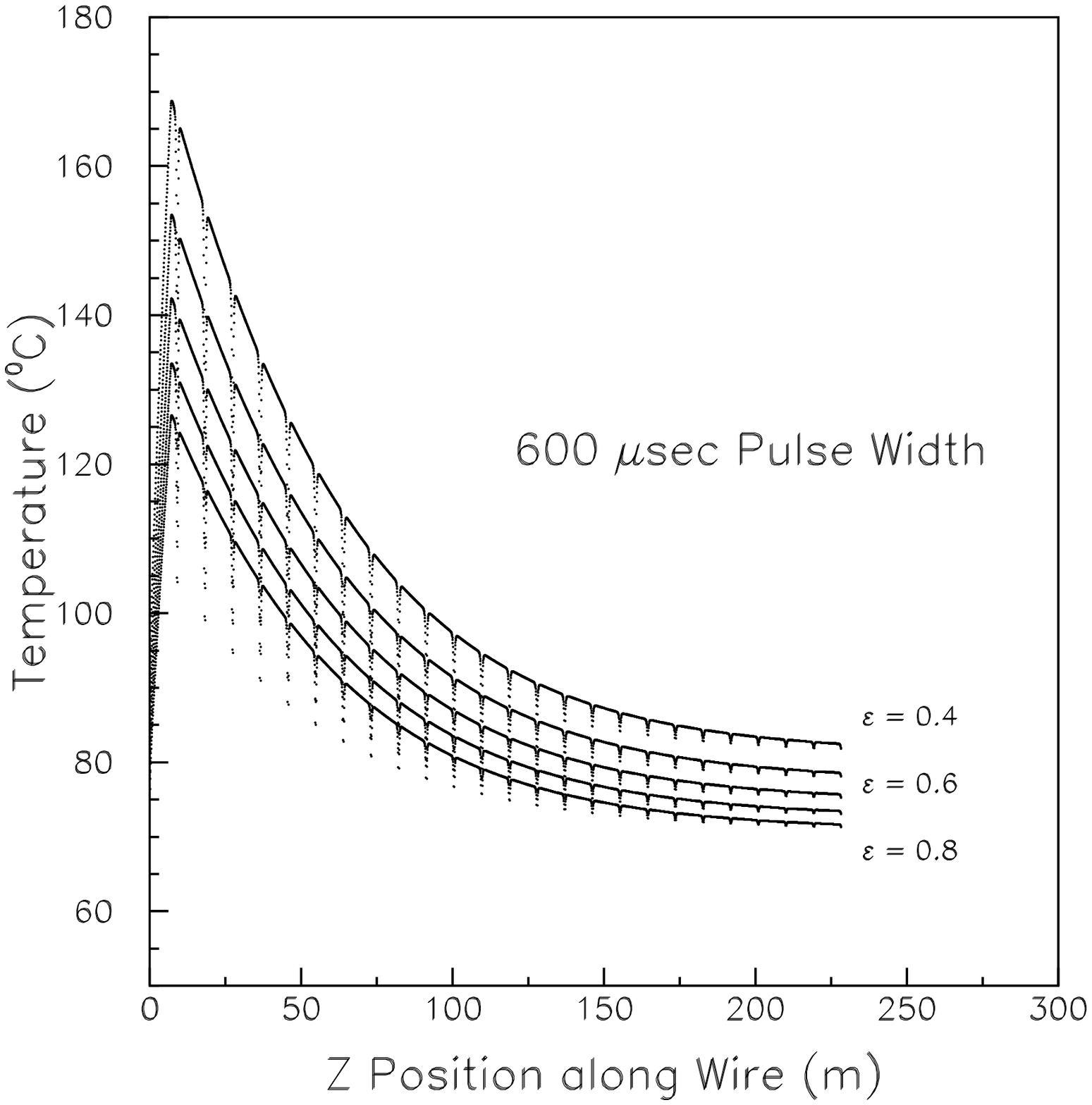}
\includegraphics[width=65mm]{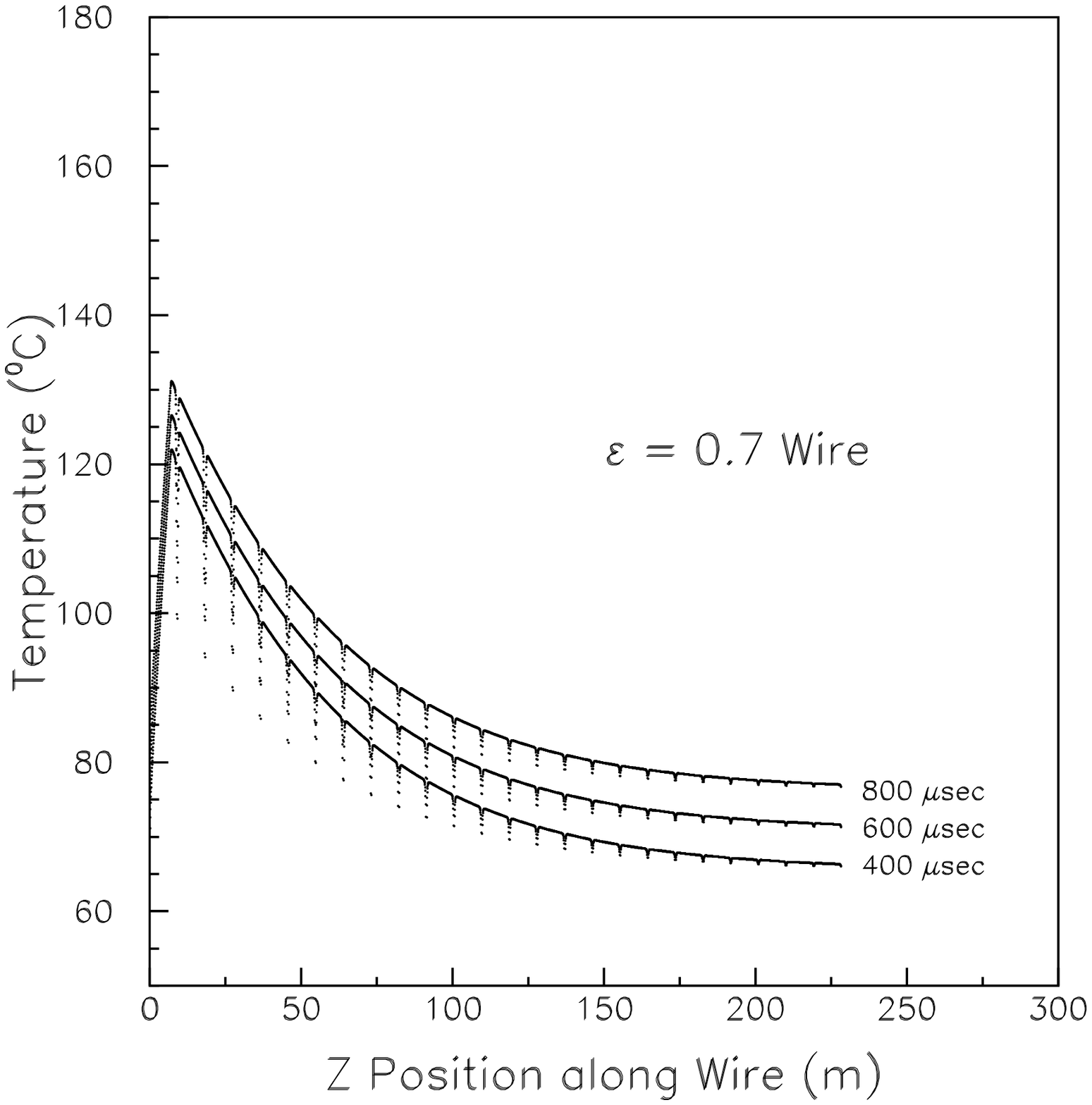}
\caption{Calculation of the hose wire temperature from a thermal model of the hose wire 
in the NuMI beam as a function of distance along the wire from the beginning of the decay volume.  (left)
Calculation for a fixed current pulse length of 600$\mu$sec half-width, several values of wire emissivity 
$\varepsilon = $0.4, 0.5 0.6, 0.7, and 0.8.  (right)  Calculation for $\varepsilon=0.7$ and current pulse
lengths of 400, 600, and 800~$\mu$sec.  }
\label{wirevsemiss}
\end{figure}

Energy deposition in the hose wire is dominated by interactions of primary protons that did not interact 
in the target.  These protons enter the decay pipe with a mean scattering angle of 0.25~mrad with respect
to the nominal beam direction.  These protons are focused by the hose field
directly into the wire.  Because the protons leave the target travelling radially, all unreacted protons 
are eventually focused into the wire.  An analytical calculation of the distance at which protons strike
the hose wire \cite{milburn} yields an energy deposition of 3.5~Watt/meter inside the first 30~m of
hose wire.  

We confirmed this calculation using the MARS beamline Monte Carlo package 
\cite{MARS}.  This simulation includes the NuMI target, focusing horns, and all material upstream of
the hose such as the target shielding and vacuum decay pipe window.  Thus, the simulation includes 
energy deposited in the wire from protons as well as by any other showering particles that 
are created in secondary interactions upstream in the beamline.  The result of this simulation, 
shown as a function of position along the hose wire, is shown in Figure~\ref{beamheating}.  
The peak at zero meters arises from protons travelling along the beam axis which interact
in the 2~m unpulsed section of the hose, while the energy deposited downstream results
from protons which multiple scatter in the target and are brought back to the beam axis by the hose field.
Not suprisingly, the simulation indicates that with the hose field turned off, this second component 
disappears.  The additional deposited energy compared to the analytic calculation indicates the magnitude
of energy deposition from showering particles created in upstream material in the beamline.

The expectation for the hose wire temperature in the NuMI beam is calculated using the thermal model
of Section~\ref{cooling}, with the addition of the beam heating and of gas cooling.  The beam heating 
term is taken from Figure~\ref{beamheating}.  The temperature of the NuMI decay pipe wall at $r_{wall}=100$~cm 
is set to $T_{wall}=55^\circ$C, based on a MARS simulation of the energy deposited in the decay pipe.
We assume a value of the heat transfer coefficient of $k = 0.0239$W/cm/$^\circ$C at 0.1-1.0~Torr 
from Figure~\ref{htrans}.
Thermal conduction is allowed to the end of each 9~m hose segment, which are also constrained to 
55$^\circ$C.  In the calculation, each hose segment is subdivided into three parts:  a central
9~m segment which receives both $i^2r$ and beam heating, and two 1~m 'leads' which bring the hose
current in from the decay pipe walls to the center of the pipe which receive only $i^2r$ heating (no beam
heating).

The results of the calculation are shown in Figure~\ref{wirevsemiss}.
In the figure, the basic shape of the temperature distribution follows the beam energy deposition
calculated in Figure~\ref{beamheating}, as may be expected. 
Based on this calculation of wire heating, a pulse length of 600~$\mu$sec is acceptable, assuming a 
wire emissivity $\varepsilon \sim 0.6-0.7$ as achieved for 17$\mu$m anodization.

\section{Radiological Issues}
\label{radiological}

The NuMI beam produces large numbers of hadrons, neutrons, gammas, and 
other
particles which can cause activation of the surrounding earth and rock.  This activation affects
underground aquifers in the vicinity of the beamline.  The NuMI decay volume and the target hall are shielded 
from the surrounding rock by poured concrete or stacked steel blocks, respectively.  The thickness of 
target hall steel is $\sim$~2~m and the decay volume concrete shield ranges from 3~m at the 
upstream end to 2~m at the downstream end.  In addition a hadron beam stop, consisting of a $2\times2$~m$^2$ by 
3~m longitudinal depth Aluminum core surrounded by a 2~m thick layer of stacked steel blocks,
is located at the end of the decay volume 

We studied whether the hose could increase the radioactivation of the surrounding earth.  Such an increase could 
result, in principle, from interactions of the remnant proton beam with the hose wire.  These interactions
occur along the full length of the decay pipe, including the most downstream sections where the shielding is
thinnest.  Without the hose, much of this component of the beam power is absorbed in the hadron stop.   

We used the MARS beamline Monte Carlo \cite{MARS} to simulate any increase in activation of the surrounding rock
resulting from such proton interactions in the wire.  All particles above 100~MeV were tracked in the simulation,
except neutrons, for which the threshold was 10~MeV.  The radioactivation measure is the
density of 'stars' in the 
surrounding rock per proton on target.  A star is a nuclear interaction above 50~MeV in the rock caused by
particles from the beam.  The density of stars is tabulated in Table~\ref{marstable}, and is averaged over
the 1~m layer of rock surrounding the shielding.  For reference, NuMI will deliver 
$4\times10^{20}$ protons on target per year.  According to the simulation, operation of the Hadron Hose does
increase the star densities in different regions around the NuMI beamline.  
These results indicate that the shielding planned for the NuMI beam line can 
accomodate inclusion of the Hadron Hose without modification.

\begin{table}[t]
\begin{center}
\begin{tabular}{|l|c|c|}
\hline
\textbf{Region} & \textbf{No Hose} & \textbf{Hose}   \\ \hline
Target Hall     & 0.59 $\pm$ 0.01  & 0.49 $\pm$ 0.06 \\
DV Upstream     & 4.4 $\pm$ 0.4    & 6.3 $\pm$ 0.4   \\
DV Middle       & 2.1 $\pm$ 0.2    & 2.4 $\pm$ 0.1   \\
DV Downstream   & 1.0 $\pm$ 0.1    & 0.9 $\pm$ 0.1   \\
\hline
\end{tabular}
\caption{Densities ($\times 10^{-11}$) of 'stars' (nuclear interactions above 50~MeV) per cm$^3$ of rock 
per proton on target in the NuMI low-energy beam.  
The star densities are tabulated for the rock surrounding the target hall and for 
three longitudinal segments of rock around the decay volume.}
\label{marstable}
\end{center}
\end{table}

\section{Conclusions}
\label{conclude}

We have investigated the potential impact of a new focusing system, the Hadron Hose, for conventional
neutrino beams.  This system was developed for the NuMI beam at Fermilab, and may be of benefit
to future conventional 'super beams' because of its increase in neutrino event yield and ability to control 
systematic uncertainties due to particle production in the target or imperfections in the rest of the 
neutrino beam elements.  

\section{Acknowledgements}
\label{acknowledgements}

We thank our NuMI and MINOS colleagues, particularly Karol Lang, Doug Michael, and Stan Wojcicki, 
for discussions and encouragement.  We thank Ron Dimelfi, John Hall, Nikolai Mokhov, Howie Pfeffer, 
Steve Sansone, and Sergei Striganov for helpful consultations.  We acknowledge the
valuable contributions of the University of Texas Physics Department 
mechanical support shops, the Fermilab Particle Physics, Beams, Survey/Alignment, 
and Mechanical Support Divisions, and in 
particular Virgil Bocean, Cary Kendziora, Stephen Pordes, Bob Webber, and James Lackey.
This work was supported by the U.S. Department of Energy, DE-AC02-76CH3000, DE-FG03-93ER40757 and 
DE-FG02-91ER40654, the National Science Foundation, NSF/PHY-0089116, and the Fondren Foundation.

\end{document}